\documentclass[9pt,twocolumn]{extarticle}
\usepackage{graphicx}
\usepackage{balance}
\usepackage{url}
\usepackage{tabularx}
\usepackage{subfigure}
\usepackage{listings}
\usepackage{amsmath}
\usepackage{amssymb}
\usepackage{amsfonts}
\usepackage{cite}
\usepackage{xspace}
\usepackage{fancyvrb,shortvrb}
\usepackage[utf8]{inputenc}
\usepackage{booktabs}
\usepackage{tikz}
\usepackage{verbatim}
\usepackage{subscript}
\usetikzlibrary{calc,trees,positioning,arrows,chains,shapes.geometric,%
    decorations.pathreplacing,decorations.pathmorphing,shapes,%
    matrix,shapes.symbols,fit,decorations.shapes}
\makeatletter
\def\PY@reset{\let\PY@it=\relax \let\PY@bf=\relax%
    \let\PY@ul=\relax \let\PY@tc=\relax%
    \let\PY@bc=\relax \let\PY@ff=\relax}
\def\PY@tok#1{\csname PY@tok@#1\endcsname}
\def\PY@toks#1+{\ifx\relax#1\empty\else%
    \PY@tok{#1}\expandafter\PY@toks\fi}
\def\PY@do#1{\PY@bc{\PY@tc{\PY@ul{%
    \PY@it{\PY@bf{\PY@ff{#1}}}}}}}
\def\PY#1#2{\PY@reset\PY@toks#1+\relax+\PY@do{#2}}

\def\PY@tok@gd{\def\PY@tc##1{\textcolor[rgb]{0.63,0.00,0.00}{##1}}}
\def\PY@tok@gu{\let\PY@bf=\textbf\def\PY@tc##1{\textcolor[rgb]{0.50,0.00,0.50}{##1}}}
\def\PY@tok@gt{\def\PY@tc##1{\textcolor[rgb]{0.00,0.25,0.82}{##1}}}
\def\PY@tok@gs{\let\PY@bf=\textbf}
\def\PY@tok@gr{\def\PY@tc##1{\textcolor[rgb]{1.00,0.00,0.00}{##1}}}
\def\PY@tok@cm{\let\PY@it=\textit\def\PY@tc##1{\textcolor[rgb]{0.25,0.50,0.50}{##1}}}
\def\PY@tok@vg{\def\PY@tc##1{\textcolor[rgb]{0.10,0.09,0.49}{##1}}}
\def\PY@tok@m{\def\PY@tc##1{\textcolor[rgb]{0.40,0.40,0.40}{##1}}}
\def\PY@tok@mh{\def\PY@tc##1{\textcolor[rgb]{0.40,0.40,0.40}{##1}}}
\def\PY@tok@go{\def\PY@tc##1{\textcolor[rgb]{0.50,0.50,0.50}{##1}}}
\def\PY@tok@ge{\let\PY@it=\textit}
\def\PY@tok@vc{\def\PY@tc##1{\textcolor[rgb]{0.10,0.09,0.49}{##1}}}
\def\PY@tok@il{\def\PY@tc##1{\textcolor[rgb]{0.40,0.40,0.40}{##1}}}
\def\PY@tok@cs{\let\PY@it=\textit\def\PY@tc##1{\textcolor[rgb]{0.25,0.50,0.50}{##1}}}
\def\PY@tok@cp{\def\PY@tc##1{\textcolor[rgb]{0.74,0.48,0.00}{##1}}}
\def\PY@tok@gi{\def\PY@tc##1{\textcolor[rgb]{0.00,0.63,0.00}{##1}}}
\def\PY@tok@gh{\let\PY@bf=\textbf\def\PY@tc##1{\textcolor[rgb]{0.00,0.00,0.50}{##1}}}
\def\PY@tok@ni{\let\PY@bf=\textbf\def\PY@tc##1{\textcolor[rgb]{0.60,0.60,0.60}{##1}}}
\def\PY@tok@nl{\def\PY@tc##1{\textcolor[rgb]{0.63,0.63,0.00}{##1}}}
\def\PY@tok@nn{\let\PY@bf=\textbf\def\PY@tc##1{\textcolor[rgb]{0.00,0.00,1.00}{##1}}}
\def\PY@tok@no{\def\PY@tc##1{\textcolor[rgb]{0.53,0.00,0.00}{##1}}}
\def\PY@tok@na{\def\PY@tc##1{\textcolor[rgb]{0.49,0.56,0.16}{##1}}}
\def\PY@tok@nb{\def\PY@tc##1{\textcolor[rgb]{0.00,0.50,0.00}{##1}}}
\def\PY@tok@nc{\let\PY@bf=\textbf\def\PY@tc##1{\textcolor[rgb]{0.00,0.00,1.00}{##1}}}
\def\PY@tok@nd{\def\PY@tc##1{\textcolor[rgb]{0.67,0.13,1.00}{##1}}}
\def\PY@tok@ne{\let\PY@bf=\textbf\def\PY@tc##1{\textcolor[rgb]{0.82,0.25,0.23}{##1}}}
\def\PY@tok@nf{\def\PY@tc##1{\textcolor[rgb]{0.00,0.00,1.00}{##1}}}
\def\PY@tok@si{\let\PY@bf=\textbf\def\PY@tc##1{\textcolor[rgb]{0.73,0.40,0.53}{##1}}}
\def\PY@tok@s2{\def\PY@tc##1{\textcolor[rgb]{0.73,0.13,0.13}{##1}}}
\def\PY@tok@vi{\def\PY@tc##1{\textcolor[rgb]{0.10,0.09,0.49}{##1}}}
\def\PY@tok@nt{\let\PY@bf=\textbf\def\PY@tc##1{\textcolor[rgb]{0.00,0.50,0.00}{##1}}}
\def\PY@tok@nv{\def\PY@tc##1{\textcolor[rgb]{0.10,0.09,0.49}{##1}}}
\def\PY@tok@s1{\def\PY@tc##1{\textcolor[rgb]{0.73,0.13,0.13}{##1}}}
\def\PY@tok@sh{\def\PY@tc##1{\textcolor[rgb]{0.73,0.13,0.13}{##1}}}
\def\PY@tok@sc{\def\PY@tc##1{\textcolor[rgb]{0.73,0.13,0.13}{##1}}}
\def\PY@tok@sx{\def\PY@tc##1{\textcolor[rgb]{0.00,0.50,0.00}{##1}}}
\def\PY@tok@bp{\def\PY@tc##1{\textcolor[rgb]{0.00,0.50,0.00}{##1}}}
\def\PY@tok@c1{\let\PY@it=\textit\def\PY@tc##1{\textcolor[rgb]{0.25,0.50,0.50}{##1}}}
\def\PY@tok@kc{\let\PY@bf=\textbf\def\PY@tc##1{\textcolor[rgb]{0.00,0.50,0.00}{##1}}}
\def\PY@tok@c{\let\PY@it=\textit\def\PY@tc##1{\textcolor[rgb]{0.25,0.50,0.50}{##1}}}
\def\PY@tok@mf{\def\PY@tc##1{\textcolor[rgb]{0.40,0.40,0.40}{##1}}}
\def\PY@tok@err{\def\PY@bc##1{\fcolorbox[rgb]{1.00,0.00,0.00}{1,1,1}{##1}}}
\def\PY@tok@kd{\let\PY@bf=\textbf\def\PY@tc##1{\textcolor[rgb]{0.00,0.50,0.00}{##1}}}
\def\PY@tok@ss{\def\PY@tc##1{\textcolor[rgb]{0.10,0.09,0.49}{##1}}}
\def\PY@tok@sr{\def\PY@tc##1{\textcolor[rgb]{0.73,0.40,0.53}{##1}}}
\def\PY@tok@mo{\def\PY@tc##1{\textcolor[rgb]{0.40,0.40,0.40}{##1}}}
\def\PY@tok@kn{\let\PY@bf=\textbf\def\PY@tc##1{\textcolor[rgb]{0.00,0.50,0.00}{##1}}}
\def\PY@tok@mi{\def\PY@tc##1{\textcolor[rgb]{0.40,0.40,0.40}{##1}}}
\def\PY@tok@gp{\let\PY@bf=\textbf\def\PY@tc##1{\textcolor[rgb]{0.00,0.00,0.50}{##1}}}
\def\PY@tok@o{\def\PY@tc##1{\textcolor[rgb]{0.40,0.40,0.40}{##1}}}
\def\PY@tok@kr{\let\PY@bf=\textbf\def\PY@tc##1{\textcolor[rgb]{0.00,0.50,0.00}{##1}}}
\def\PY@tok@s{\def\PY@tc##1{\textcolor[rgb]{0.73,0.13,0.13}{##1}}}
\def\PY@tok@kp{\def\PY@tc##1{\textcolor[rgb]{0.00,0.50,0.00}{##1}}}
\def\PY@tok@w{\def\PY@tc##1{\textcolor[rgb]{0.73,0.73,0.73}{##1}}}
\def\PY@tok@kt{\def\PY@tc##1{\textcolor[rgb]{0.69,0.00,0.25}{##1}}}
\def\PY@tok@ow{\let\PY@bf=\textbf\def\PY@tc##1{\textcolor[rgb]{0.67,0.13,1.00}{##1}}}
\def\PY@tok@sb{\def\PY@tc##1{\textcolor[rgb]{0.73,0.13,0.13}{##1}}}
\def\PY@tok@k{\let\PY@bf=\textbf\def\PY@tc##1{\textcolor[rgb]{0.00,0.50,0.00}{##1}}}
\def\PY@tok@se{\let\PY@bf=\textbf\def\PY@tc##1{\textcolor[rgb]{0.73,0.40,0.13}{##1}}}
\def\PY@tok@sd{\let\PY@it=\textit\def\PY@tc##1{\textcolor[rgb]{0.73,0.13,0.13}{##1}}}

\def\PYZus{\char`\_}
\def\PYZob{\char`\{}
\def\PYZcb{\char`\}}

\def\PYZpc{\char`\%}

\makeatother

\definecolor{orange}{RGB}{205,102,0}

\newcommand{\optimizer}{\textsc{Sofa}\xspace}
\newcommand{\ontology}{Presto\xspace}
\newcommand{\stratosphere}{Stratosphere\xspace}
\newcommand{\udfs}{\textsc{Udf}s\xspace}
\newcommand{\udf}{\textsc{Udf}\xspace}

\newcommand{\ie}{i.\,e.\@\xspace}

\newcommand{\etal}{et~al.\@\xspace}

\newcounter{foo}
\newenvironment{myenumerate}{\begin{list}{\arabic{foo}.}{
        \usecounter{foo}
        \setlength{\leftmargin}{15pt}
        \setlength{\topsep}{2pt}
       \setlength{\parsep}{0pt}
        \setlength{\itemsep}{1pt}
}}{ \end{list}}

\newenvironment{myitemize}{\begin{list}{--}{
        \setlength{\leftmargin}{15pt}
        \setlength{\topsep}{2pt}
        \setlength{\parsep}{0pt}
        \setlength{\itemsep}{1pt}
}}{ \end{list}}

\title{SOFA: An Extensible Logical Optimizer for UDF-heavy Dataflows}

\author{
    Astrid Rheinl\"ander\\
    Humboldt-Universit{\"a}t zu Berlin\\
    Berlin, Germany\\
    rheinlae@informatik.hu-berlin.de
  \and
    Arvid Heise\\
    Hasso Plattner Institut\\
    Potsdam, Germany\\
    arvid.heise@hpi.uni-potsdam.de
    \and
    Fabian Hueske\\
    Technische Universit\"at Berlin\\
    Berlin, Germany\\
    fabian.hueske@tu-berlin.de
    \and
    Ulf Leser\\
    Humboldt-Universit{\"a}t zu Berlin\\
    Berlin, Germany\\
    leser@informatik.hu-berlin.de
    \and
    Felix Naumann\\
    Hasso Plattner Institut\\
    Potsdam, Germany\\
    felix.naumann@hpi.uni-potsdam.de
}

\date{\vspace{-4ex}}

\begin{document}
\maketitle

\begin{abstract}
Recent years have seen an increased interest in large-scale analytical dataflows on non-relational data.
These dataflows are compiled into execution graphs scheduled on large compute clusters.
In many novel application areas the predominant building blocks of such dataflows are user-defined predicates or functions (\udfs).
However, the heavy use of \udfs is not well taken into account for dataflow optimization in current systems.

\optimizer is a novel and extensible optimizer for \udf-heavy dataflows.
It builds on a concise set of properties for describing the semantics of Map/Reduce-style \udfs and a small set of rewrite rules, which use these properties to find a much larger number of semantically equivalent plan rewrites than possible with traditional techniques.
A salient feature of our approach is extensibility:
We arrange user-defined operators and their properties into a subsumption hierarchy, which considerably eases integration and optimization of new operators.
We evaluate \optimizer on a selection of \udf-heavy dataflows from different domains and compare its performance to three other algorithms for dataflow optimization.
Our experiments reveal that \optimizer finds efficient plans,
outperforming the best plans found by its competitors by a factor of up to 6.
\end{abstract}

\section{Introduction}\label{sec:introduction}

In recent years, the characteristics of data analysis tasks have changed significantly.
One change is the increase in the typical amounts of data to be processed; in addition, the diversity of the data to be analyzed and the heterogeneity of analysis tasks has grown considerably. 
While in the past most analytics were performed on structured data using relational data processors, such as SQL OLAP, many applications today require complex analyses, such as heavy-weight machine learning, predictive modeling, or graph traversal on large texts, graphs, semi-structured datasets, etc.
Data processing systems support such analyses by providing system APIs for user-defined functions (\udf).
Commonly, these \udfs are specified with imperative code, registered with the system, and called during execution.
In fact, a large portion of Map/Reduce's popularity can be accounted to its widespread support for custom data processing~\cite{dean_mapreduce_2004}.

A variety of dataflow languages has been proposed that aim at (a)~making the definition of complex analytics tasks easier and at (b)~allowing flexible deployment of such dataflows on diverse hardware infrastructures, especially compute clusters or compute clouds~\cite{sakr_survey_2011}.
Many of these languages support \udfs~\cite{heise_meteor_2012,olston2008pig,beyer2011jaql}.
Research has shown that proper optimization of such dataflows can improve the execution times by orders of magnitude~\cite{cafarella2010manimal,hueske_opening_2012,wu_query_2011}.
However, most of these systems focus on relational operators and treat \udfs essentially as black boxes, considerably hampering optimization.
At the same time, non-standard applications in areas such as information extraction, graph analysis, or predictive modeling often utilize \udf-rich dataflows. Traditional optimizers focus on relational operations, because their semantics in terms of optimization are well understood. In contrast, Map/Reduce-style \udfs can exhibit all sorts of behavior, which are difficult to describe in an abstract, optimizer-enabling manner.

Different approaches have been proposed to overcome this problem.
Two broad classes can be discerned: Approaches that require manual annotations of \udfs~\cite{lim2012stubby, battre_nephele_2010}, and approaches performing code analysis to automatically infer optimization options with \udfs~\cite{cafarella2010manimal,guo_spotting_2012,hueske_opening_2012}.
We present \optimizer, a semantics-aware optimizer for \udf-heavy dataflows\footnote{\optimizer is only a vague acronym but more of a metaphor.}. Compared to previous work, \optimizer features a richer, yet concise set of properties for automatic and manual \udf annotation.
It is capable of finding a much larger and more efficient set of semantically equivalent execution plans for a given dataflow than other systems. Given a concrete dataflow, both automatically detected and manually created annotations are evaluated by a cost-based optimizer, which uses a concise set of rewrite templates to infer semantically equivalent execution plans.
\begin{figure*}
\centering
\includegraphics[width=1\linewidth]{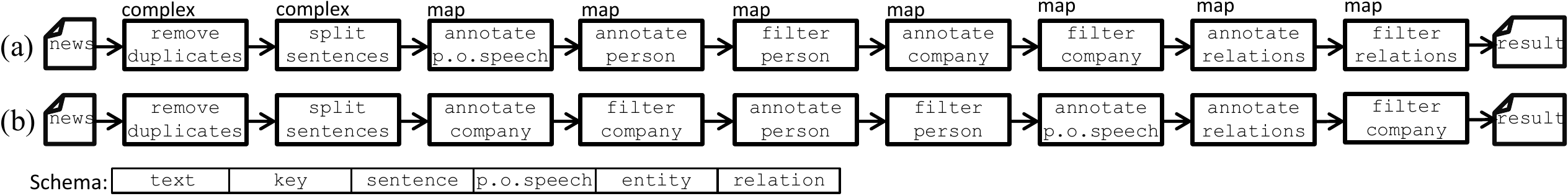}
\vspace{-.5cm}
\caption{High-level dataflow for employee relationship analysis.  (a)~Initial dataflow, (b)~reordered dataflow based on operator semantics.}
\label{fig:exintro}
\end{figure*}

\vspace{.1cm}
\noindent\textbf{Example.} We use the following \emph{running example} to explain the principles of \optimizer throughout this paper:
A large set of news articles shall be analyzed to identify persons, companies, and
associations of persons to companies.
We assume the articles stem from a web crawl and have already been stripped of HTML tags, advertisements, etc.; still the set contains many duplicate articles, as different news articles are often copied from  reports prepared by a news agency.

~

An exemplary dataflow for this task is shown in Figure~\ref{fig:exintro}(a).
The first operator performs duplicate removal by first computing a grouping key followed by an analysis of each group for similar documents, such that detected duplicates are filtered out per group.
Next, a series of operators performs linguistic analysis (sentence splitting and part-of-speech tagging), entity recognition (persons and companies), and relation identification (persons $\leftrightarrow$ companies).
After each annotation operator, filter operators remove texts with no person, no company, or no person-company relation, respectively.
As displayed in Figure~\ref{fig:exintro}(a), the dataflow is composed of nine steps: seven maps, and two complex operators (first and second from left).

If \udfs are treated as black boxes, this dataflow cannot be reordered in any way.
But when provisioned with proper information, such as data dependencies or operator commutativities, an optimizer has multiple options for reordering.
For example, the part-of-speech tagger can be pushed multiple steps toward the end of the dataflow, as annotations produced by this operator are necessary for relationship annotation only.
Moreover, the company and person annotation operators are commutative, as they independently add annotations to the text, but never delete existing annotations.
Thus, both annotation operators can be reordered for early filtering. 
Figure~\ref{fig:exintro}(b) displays an equivalent dataflow with prospectively smaller costs as the most selective filters are executed as early as possible and expensive predicates are moved as much to the end of the dataflow as possible. 
As we see in Sections~\ref{sec:semopt} and~\ref{sec:evaluation}, existing dataflow optimization techniques, such as~\cite{hueske_opening_2012,olston2008pig}, cannot infer this plan.

A major obstacle to the optimization of Map/Reduce-style \udfs is their diversity.
Our algorithm is developed in \stratosphere, a platform for data analytics on massive datasets with custom,
domain-specific operator packages~\cite{battre_nephele_2010}.
Available packages for information extraction (IE) and data cleansing (DC) already contain 38 and 9 operators, respectively. Further packages, e.g., for machine-learning and web analytics, are in development. Defining semantic properties for each of these operators individually would result in an unacceptable burden to the designer and would considerably limit extensibility and maintainability. Furthermore, the optimizer needs rewrite rules for operator pairs that take operator semantics into account. Thus, every new operator in principle has to be analyzed with respect to every existing operator to specify possible rewritings. \optimizer solves this problem by means of an extensible \emph{taxonomy} of operators, operator properties, and rewrite templates called \textit{Presto}. \optimizer uses the information encoded in \ontology to reason about relationships between operators during plan optimization; specifically, it leverages subsumption relationships between operators to derive reorderings not explicitly modeled.
\ontology considerably eases extensions, as novel operators can be hooked to the system by specifying a single subsumption relationship to an existing operator exhibiting the same behavior with respect to optimization; these new operators are immediately optimized in the same manner as their parent.
If desired and appropriate, more rewrite rules describing specific properties of the new operator may be introduced later in a ``pay-as-you-go'' manner~\cite{roth_dont_1997}.

\vspace{-3mm}~\\
\noindent In summary, our work includes the following contributions:
\begin{myenumerate}
\item We identify a small yet powerful set of common \udf properties influencing important aspects in terms of dataflow optimization.
\item We show how properties of \udfs in Map/Reduce-style systems can be arranged in a concise
taxonomy to simplify \udf annotation, and to enhance extensibility of dataflow languages.
\item We present a novel optimization algorithm that is capable of rewriting DAG-shaped dataflows given proper operator annotations.
\item We evaluate our approach using a diverse set of dataflows cutting through different domains. We show that \optimizer subsumes three existing dataflow optimizers by enumerating a larger plan space and by finding more efficient plans.
\item \optimizer outperforms the best plans found with other approaches in many situations with factors of up to~6.
\end{myenumerate}

\noindent This paper is structured as follows:
Section~\ref{sec:background} describes preliminaries including a brief overview of \stratosphere with a focus on its rich set of domain-specific \udfs.
Section~\ref{sec:semopt} gives an overview of our approach for dataflow optimization.
Details on \ontology and \optimizer are explained in  Sections~\ref{sec:approach} and~\ref{sec:algorithms}.
Section~\ref{sec:relatedwork} discusses related work.
We evaluate our approach in Section~\ref{sec:evaluation} and conclude in Section~\ref{sec:conclusions}.

\section{Preliminaries}
\label{sec:background}

We study the optimization of deterministic dataflows  on a semi-structured data model, such as JSON or XML\@. A \textit{record} is a potentially nested value tree consisting of objects, arrays, and atomic values. An unordered bag of records is called \textit{dataset}.
An \emph{operator} $o$ transforms a list of input datasets $I=(I_1,\ldots,I_n)$ into a list of output datasets $O=(O_1,\ldots,O_m)$ by applying a \udf $f$ to $I$.
We denote with $o_{in,i}$ and $o_{out,i}$ the $i$-th input and output of operator~$o$.
We call $S(o_{in,i})$ \emph{input schema} and $S(o_{out,i})$ \emph{output schema} of the $i$-th input and output of $o$, respectively.

A \emph{dataflow} is a connected directed acyclic graph $D=(V,E)$ with the following properties: Vertices in $D$ are either operators, data sources, or data sinks.
Sources have no incoming and sinks no outgoing edges, respectively.
Inner nodes with both incoming and outgoing edges are concrete operators.
For any directed edge $(u,v)$ connecting two operators we demand $S(u_{out,j})\supseteq S(v_{in,i})$, and $S(v_{in,i})$ must meet the schema requirements of the $i$-th input of $v$; according conditions hold for edges from an input dataset or to a data sink.
Our data model does not require a schema definition in the first place, but operators might require that the processed records have a certain schema.

We call two deterministic dataflows $D, D'$ \textit{semantically equivalent} (denoted $D \equiv D'$), if $D$ and $D'$ produce the same output sets $T$, given the same input datasets $I$.

\subsection{\stratosphere}
\begin{figure}[t]
\centering
\includegraphics[width=.75\linewidth]{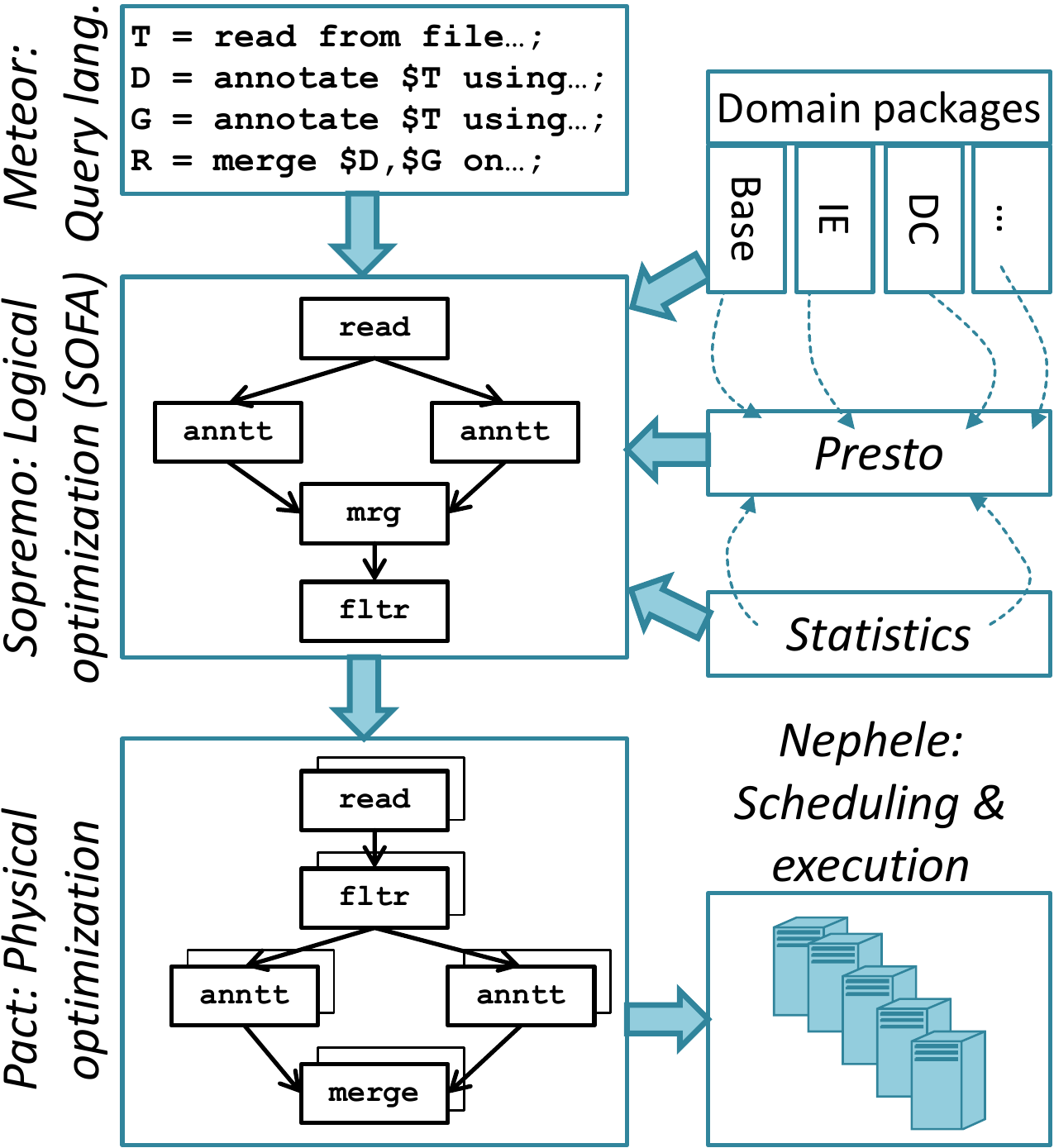}
\caption{Architecture of \stratosphere.}
\label{fig:stratosphere_stack}
\end{figure}

The \optimizer optimizer is integrated into \stratosphere, a full-fledged system for massively parallel data analytics.
Meteor~\cite{heise_meteor_2012}, a dataflow-oriented declarative scripting language resides at
the top of the stack (see Figure~\ref{fig:stratosphere_stack}).
A Meteor query is parsed into an abstract syntax tree composed of basic or complex operators (see
Section~\ref{sec:semopt}) and then compiled into a logical execution plan in the system's algebra called Sopremo. \optimizer resides on the algebraic
layer of \stratosphere and employs information on properties and semantics of operators stored in the \ontology taxonomy (see Section~\ref{sec:approach}) as well as statistics on the operators to perform dataflow optimization (see Section~\ref{sec:algorithms}).

Meteor and Sopremo are designed for extensibility.
Operators are defined in domain-specific packages, which are self-contained libraries of the operator implementations, their syntax, and semantic annotations. 
Algebraic Sopremo plans are compiled into Pact programs, which may consist of different parallelization primitives,
such as map or reduce, and the associated user-defined function~\cite{battre_nephele_2010}.
Subsequently, Pact programs are physically optimized and translated into an execution graph, which is deployed on the
given hardware by means of Nephele, a system for scheduling, executing, and monitoring DAG structured execution graphs
on distributed systems. A detailed description of \stratosphere can be found in~\cite{battre_nephele_2010}. Here, we describe only the algebraic optimizer.

\newcolumntype{C}[1]{>{\centering}m{#1}}
\begin{table}[t]
\centering
\scriptsize
\begin{tabular}{llccc}
\toprule
Operator		 	& Abbre-					&\#in-   	& Process-	& Preserves \\
 					& viation					& puts	& ing type 		& schema? \\
\midrule
filter 				& \textit{fltr}			& 1 	& Record 	& Yes 	\\
project 			& \textit{prjt}			& 1 	& Record 	& No	\\
(un)nest			& \textit{nst}			& 1 	& Record 	& No	\\
join 				& \textit{join}			& n 	& Record 	& No	\\
(co)group			& \textit{grp}			& 1/2 	& Bag 		& No	\\
\ldots				&					&	&		&	\\
\midrule
annotate entities 		& \textit{anntt-ent}		& 1 	& Record 	& Yes 	\\
annotate tokens			& \textit{anntt-tok}		& 1 	& Record 	& Yes 	\\
merge 				& \textit{mrg}			& 2 	& Bag 		& Yes 	\\
split sentences 		& \textit{splt-sent}		& 1 	& Record 	& Yes 	\\
extract relations 		& \textit{extr-rel}			& 1 	& Record 	& No 	\\
remove stopwords 		& \textit{rm-stop}			& 1 	& Record 	& Yes 	\\
\ldots				&					&	&		&	\\
\midrule
scrub  				& \textit{scrb}			& 1 	& Bag	 	& Yes	\\
split records			& \textit{sptrc}			& 1 	& Record 	& No	\\
transform records		& \textit{trfrc}			& 1 	& Record 	& No	\\
detect duplicates		& \textit{ddup}			& 1 	& Bag	 	& No	\\
remove duplicates		& \textit{rdup}			& 1 	& Bag	 	& No	\\
fuse 				& \textit{fuse}			& 1 	& Record 	& No	\\
\ldots				&					&	&		&	\\
\bottomrule
\end{tabular}
\caption{Selected general purpose (top), IE (middle), and DC (bottom) operators available for Stratosphere. See~\cite{heise_meteor_2012} for details.}
\label{tab:packages}
\end{table}

\subsection{User-Defined Operators}
Before we introduce our \udf annotation schema and dataflow optimizer, we first introduce a subset of the operators
currently used in \stratosphere to highlight their diversity and resulting optimization challenges. User-defined
operators are organized into \textit{packages} specific to a certain application domain.
Operators can be either abstract or concrete; for instance, \textit{anntt-ent-pers} (see Table~1) is an abstract operator for annotating person names in texts, and its instantiations are different concrete algorithms and tools for performing this task.
All instantiations of a given abstract operator share the same language syntax.
Note that concrete operators may use very different implementations; for instance, the recognition of person names may be performed by using dictionary-, pattern-, or machine-learning-based methods.

Concrete operators can either be elementary or complex.
Elementary operators are implemented using a single stub function, complex operators are composed of several elementary operators similar to macros in programming languages.
Complex operators are of high practical relevance, as they provide a shortcut for adding multiple elementary operators to a query.
They are also important for dataflow optimization, since a complex operator may exhibit other semantics than those of its elements (see below).

\stratosphere currently contains three packages, namely a base package, a package for IE, and a package for DC\@.
Packages for machine-learning and web data analytics are under development.
Table~\ref{tab:packages} shows example operators from all three packages together with information describing their requirements and behavior.
The base package contains 16 operators, the IE package 38 operators, and the DC package 9 operators.
The base package mostly comprises typical relational operators, such as filter, projection, transformation, join, and group. These operators are complemented by operators for semi-structured data, such as nest or unnest.

The IE package comprises three classes of operators: One for producing text annotations, one to merge annotations, and one for complex operators.
Operators analyze the text and add, remove, or update annotations to the record.
They may also transform records, e.g., the \textit{splt-sent} operator takes as input single records formed of documents and outputs a set of records formed of sentences.
The most abstract operator in the annotation class is \textit{anntt}.
Specializations can be distinguished between those performing linguistic annotations, semantic annotation of entities, or semantic annotations of relationships between entities.
Each of these classes consists of multiple concrete operators; e.g., operators for tokenization, or part-of-speech tagging (first group), for recognizing persons, companies, or biomedical entities (second group), and for detecting binary or n-ary relationships between entities (third group).
Specializations of \textit{anntt} write to designated attributes in the output record; for instance, all entity operators write to a list-valued field ``entities".
Some \textit{anntt} specializations are in precedence relations with other \textit{anntt} variants, for example, annotating relations between entities requires that entity annotations are already present in the input records.
The merge operator \textit{mrg} merges records $a,b$ from two input sets $A,B$ based on a user-defined merge condition.
The set of complex operators comprises six operators, such as operators for splitting text into sentences, stemming, entity extraction, and stopword removal.
Each of these internally consists of an \textit{anntt} operator, a \textit{trnsf} operator, and occasionally a \textit{fltr} operator.

The DC package comprises six different classes of operators for data cleansing and data integration~\cite{StratosphereIntegration}
They address common challenges of dirty or heterogeneous data sources, such as inconsistent representation of equivalent values, fuzzy duplicates, typographic errors, or missing values.
Inconsistencies and missing values can be fixed with the \textit{scrb} operator that either repairs these values or filters invalid records.
Fuzzy duplicates are found with \textit{ddup} and \textit{lnkrc} within one relation or across several relations, respectively.
These duplicates can subsequently be coalesced into a single record with \textit{fuse}.
The complex duplicate removal operator (\textit{rdup}) combines duplicate detection and fusion of duplicates to solve the common task of removing all duplicate records within one data source.

\begin{figure*}
\centering
\includegraphics[width=0.9\linewidth]{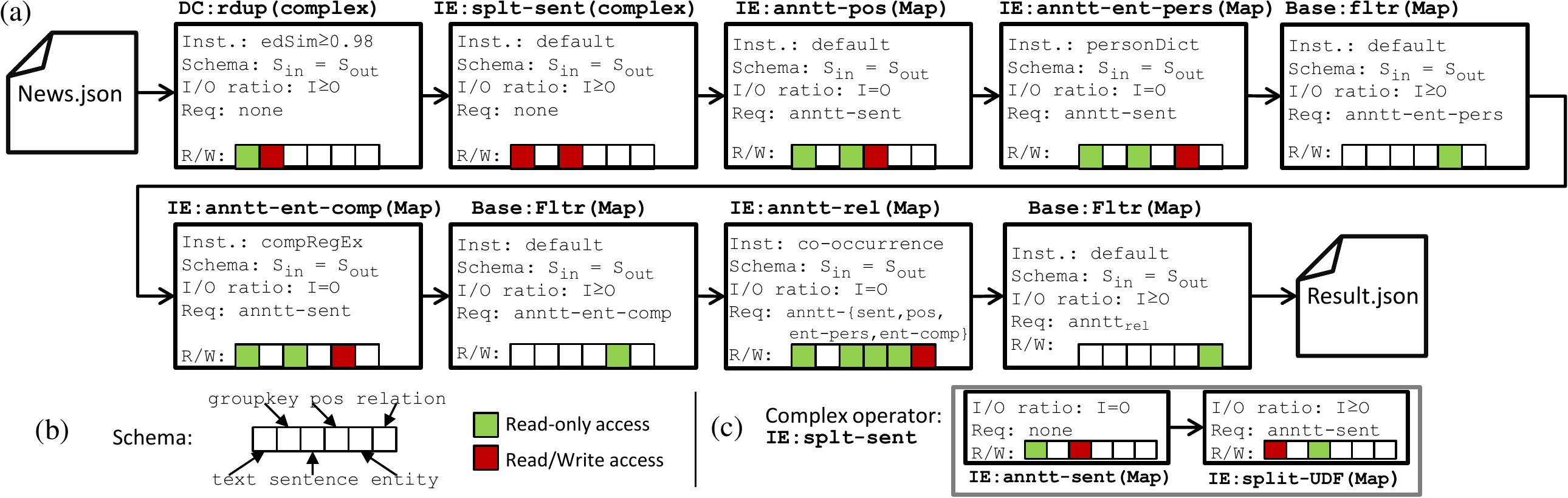}
\caption{Algebraic dataflow for the running example.  (a) displays concrete operator instantiations together with properties relevant
for optimization. Colored boxes indicate read/write access on
record attributes, which are part of the global schema shown in (b). (c) shows the resolution of the complex
\textit{splt\textsubscript{sent}} operator (second operator in (a)) into its components \textit{anntt\textsubscript{sent}} and \textit{split-UDF}.}
\label{fig:exampledf}
\end{figure*}

The algebraic plan for the dataflow for our running example is shown in Figure~\ref{fig:exampledf}(a) together with properties and schema information (cf.\ Figure~\ref{fig:exampledf}(b)), which are used for optimization. Figure~\ref{fig:exampledf}(c) displays that complex operators may exhibit different properties than its elementary components: the complex operator \textit{splt-sent} has different read/write set annotations and different I/O ratios than its elementary components. In the following section, we demonstrate how this plan can be reordered substantially by exploiting information on operator semantics.

\section{Dataflow Optimization}
\label{sec:semopt}

\optimizer is an optimizer for rewriting \udf-heavy dataflows into semantically equivalent dataflows whose expected efficiency is higher, according to a cost model.
Rewriting depends on a set of rewrite rules, each defining valid manipulations of dataflow sub-plans, such as a reordering of two filter operations~\cite{graefe_volcano_1994}.
The novelty of \optimizer is its flexible and extensible treatment of Map/Reduce-style \udfs going far beyond the capabilities of existing approaches.
In this section, we highlight the advantages of \optimizer by means of our running example.

Starting from a dataflow $D$, dataflows semantically equivalent to $D$ may be produced using different transformation techniques. \optimizer is capable of introducing, removing, and reordering operators. Complex operators are optimized both as a whole and after expansion.
In the following, we focus on reordering and treatment of complex operators, which are the most intricate and most effective optimization techniques.

Existing approaches for dataflow optimization enable reorderings by using either manually defined rewrite rules for relational operators~\cite{olston2008pig} or by performing some kind of code analysis~\cite{hueske_opening_2012,cafarella2010manimal}.
The approach of Hueske et al.~\cite{hueske_opening_2012} probably is the most general, as it automatically derives dataflow reorderings based on read/write set analysis of individual \udfs.
In particular, the order of two subsequent tuple-at-a-time operators may safely be switched if they have no read/write or write/write conflicts on any attribute.
The dataflow shown in Figure~\ref{fig:exampledf} allows only one such reordering: The \textit{anntt-pos}
operator that annotates part-of-speech tags and stores them in the fourth attribute (first row, third from right) can be pushed
before the \textit{anntt-rel} operator (second row, second from right), because part-of-speech annotations are accessed only during relation annotation.
This reordering most likely saves time, because the different \textit{fltr} operators are now executed before \textit{anntt-pos} and thus fewer sentences have to be annotated.
Moving \textit{anntt-pos} towards the start of the dataflow is not possible, because it reads annotations produced by the complex \textit{splt-sent} operator.

Semantics-aware rewrite rules allow to reorder the dataflow in Figure~\ref{fig:exampledf} more extensively.
Consider the two \textit{anntt-ent} operators.
Both write into the same attribute (the fifth), which collects all entity information.
If the optimizer knows that annotation operators only \emph{add} values and never delete or update existing values, these operators together with their subsequent filter operators may be reordered.
The best order very likely is the one that filters the most sentences first; this decision can make use of selectivity and execution time estimates at the operator level (see Section~5.3). 
Furthermore, the optimizer can decompose complex operators and reorder the components individually.
For instance, \textit{splt-sent} consists of an \textit{anntt-sent} operator and a \udf splitting the input text into separate sentences based on the annotations produced by \textit{anntt-sent}.
As shown in Figure~\ref{fig:exampledf}(c), the two components of \textit{splt-sent} differ in terms of read and write access on attributes and I/O ratio.
Pushing split-\udf some steps towards the end of the plan is valid, because all succeeding \textit{anntt} operators
perform their ana\-lyses sentence-based, since all \textit{anntt} operators for entity, relation, and part-of-speech
annotation read sentence annotations.
This reordering is likely beneficial, because the split-\udf generates multiple output records for each incoming record, depending on the number of annotated sentence boundaries.

Note that sometimes the expansion of complex operators may make possible reorderings impossible. For instance, think of an IE operator \textit{anntt-syns} for expanding existing entity annotations with all synonyms from a dictionary.
Subsequently, another operator \textit{repl-repr} selects one of these synonyms as representative and deletes all others.
Both operators change the existing list of entity annotations, so reordering based on read/write-set analysis is limited.
However, a complex operator \textit{norm-ent} for normalizing entity annotations composed of both \textit{anntt-syns} and \textit{repl-repr} is optimizable, since we know that the number of entities does not change. Therefore, \optimizer always tries to reorder complex operators both as a whole \emph{and} after their expansion.

In summary, semantics-aware plan rewriting allows us to pick the best plan (with respect to cost estimates) from a larger set of equivalent dataflows compared to other existing approaches.
For instance, \optimizer finds 4,545 distinct plans for the running example, compared to only 512 plans found with the read/write-set analysis of~\cite{hueske_opening_2012} (see Section~\ref{sec:evaluation} for a detailed comparison).

\begin{figure*}
  \centering
  \subfigure[Operator taxonomy]{
    \label{subfig:optax}
    \includegraphics[width=0.35\linewidth]{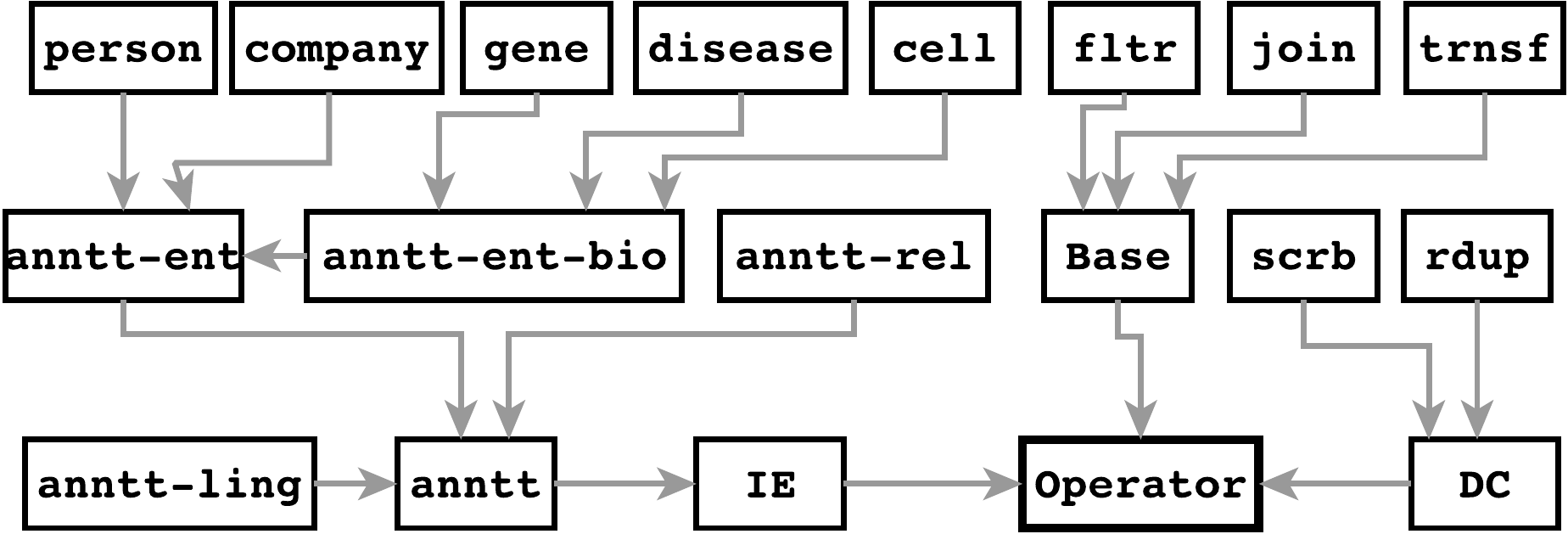}
    \setcounter{subfigure}{1}
  }
  \subfigure[Property taxonomy]{
    \label{subfig:proptax}
    \includegraphics[width=0.49\linewidth]{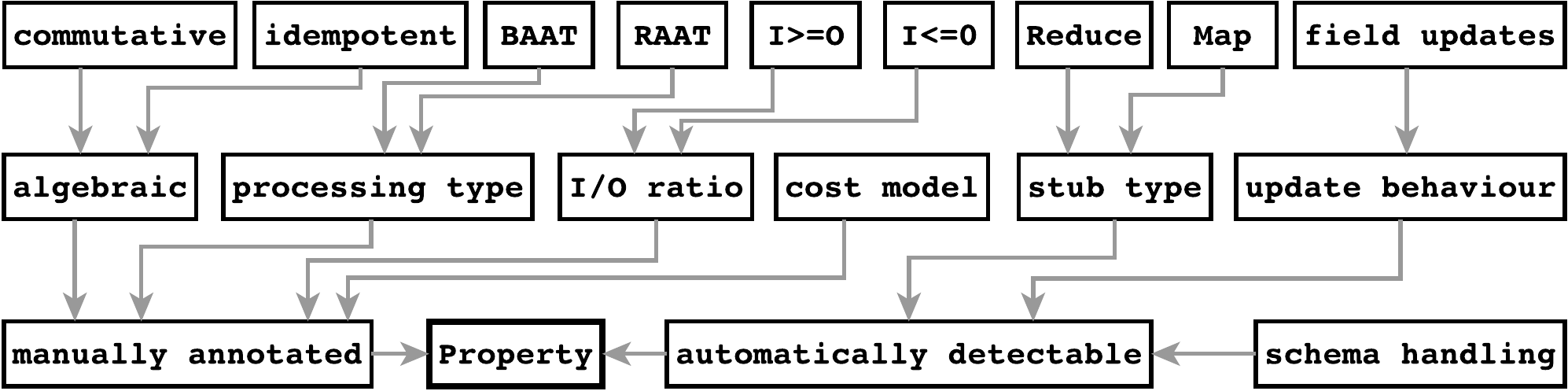}
    \setcounter{subfigure}{2}
  }\\
   \subfigure[hasProperty]{
    \label{subfig:optax}
    \includegraphics[width=0.50\linewidth]{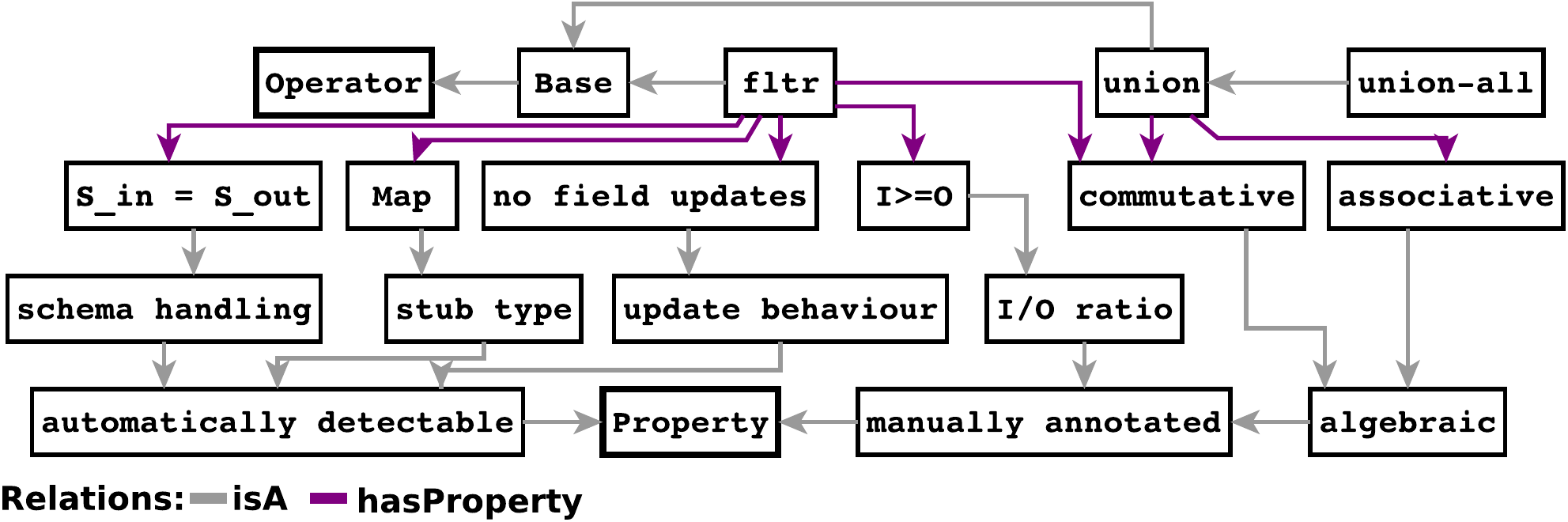}
    \setcounter{subfigure}{3}
  }
  \qquad
  \subfigure[hasPrerequisite/hasPart]{
    \label{subfig:proptax}
    \includegraphics[width=0.26\linewidth]{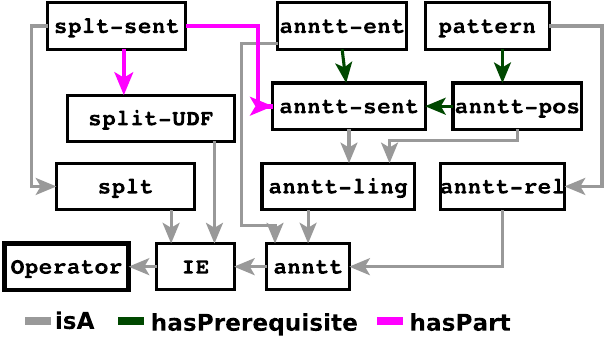}
    \setcounter{subfigure}{4}
  }
  \caption{Exemplary subgraphs of \ontology; root nodes are displayed in bold.}
  \label{fig:presto}
\end{figure*}

\section{Annotating and Rewriting \\ Operators with Presto}
\label{sec:approach}

To enable optimizations like those shown in the previous section, operators need to be annotated with meta data, for instance to describe selectivity estimates or semantic properties, such as associativities or commutativities. In this section, we introduce \ontology, an extensible taxonomy for annotating and rewriting operators. \ontology consists of two major components, the operator-property graph for modeling relationships between operators and properties (see Section~4.1), and a set of rewrite templates for dataflow-rewriting (see Section~4.2).
When designing \ontology, we paid special attention to extensibility by allowing enhancements to the semantic operator descriptions  over time, to more and more unleash their optimization potential (see Section~4.3).

\subsection{Operator-Property Graph}
The operator-property graph in \ontology contains two taxonomies for classifying \emph{operator properties}.
Both taxonomies are self-contained and model generalization--spe\-cia\-li\-zation relationships (\textit{isA}) between operators and properties, respectively.
Figures~\ref{fig:presto}(a) and (b) display subgraphs of \ontology.
For example, the \textit{anntt} operator has two specializations: \textit{anntt-ent} and \textit{anntt-rel} shown in Figure~\ref{fig:presto}(a). 
Each leaf in the operator taxonomy describes a concrete implementation of the parent operator, like different implementations for a join. This design allows us to uniquely identify available operator instantiations, to use subsumption to effectively assign properties and relationships to operators, and to deduce rewrite options. Note that such abstraction-implementation relationships are an established concept in relational optimizers. 
However, in the relational world the hierarchies are very flat; they become much deeper when dealing with domain-specific \udfs.

As shown in Figure~\ref{fig:presto}(b), we distinguish between automatically detectable properties and properties that are annotated by the package developer.
The latter comprises algebraic properties (e.g., commutativity, associativity, or idempotence), cost model (e.g., cost functions, resource consumption), and the ratio between the number of input and output records.
Automatically detectable properties comprise the parallelization function of the operator implementation (e.g., map, reduce), schema information available at query compile time, and the read/write behavior at attribute level.

Relationships connect operators and properties. As usual, each specialization inherits all properties and relationships that are defined for the corresponding generalizations.
For instance, the \textit{union-all} shown in Figure~\ref{fig:presto}(c) is a specialization of the \textit{union}
operator and thus inherits the algebraic properties defined for \textit{union}. Complex operators can be characterized with respect to their components using the \textit{hasPart} relation (Figure~\ref{fig:presto}(d)). For example, the complex operator \textit{splt-sent} consists of the two components \textit{anntt-sent} and \textit{split-UDF}.
Next to \textit{isA} and \textit{hasPart}, we define a \textit{hasProperty} and a \textit{hasPrerequisite} relation.

\textit{HasProperty} is a binary relation between an operator and a property and is used to characterize operator semantics. For instance, the following properties are attached to \textit{fltr} (Figure~\ref{fig:presto}(c)):
\begin{myitemize}
\item does not modify the schema of incoming records,
\item is implemented with a Map function,
\item does not modify inside fields,
\item input $\ge$ output,
\item processes one record at a time, and
\item is commutative to other \textit{fltr} instantiations.
\end{myitemize}

Precedence constraints between operators are captured with \textit{hasPrerequisite(X,Y)}, which states that operator $X$ must be executed before operator $Y$.
In Figure~\ref{fig:presto}(d) it is shown that \textit{anntt-rel} based on linguistic patterns requires part-of-speech and entity annotations to be performed in advance.
Since \textit{anntt-ent} itself requires sentence annotation and \textit{hasPrerequisite} is a transitive relation, it is necessary to apply \textit{anntt-sent} before \textit{anntt-rel}.

The \textit{isA} relationship very much simplifies deriving novel rewrite options for operators that are initially not well annotated. Suppose, the data scrubbing operator \textit{scrb} from the DC package is initially not equipped with any \textit{hasProperty} relationships. Later, the developer may see that \textit{scrb} is a specialization of the well-annotated Base operator \textit{trnsf}, i.e., both operators perform write operations in attributes of the incoming records. By formally specifying this through an \textit{isA} relationship, \textit{scrb} inherits all properties defined for \textit{trnsf} (not shown in Figure~\ref{fig:presto}(a)).

Though the complete \ontology graph is too large to show here, it is still rather small and easy to understand: The property taxonomy contains 32 nodes and the operator taxonomy 78 nodes.
Note that new packages mostly extend the operator taxonomy, while the property taxonomy is a fairly stable structure in our experience.

\subsection{Rewrite Templates}
We perform dataflow-rewriting using a set of rules specifying semantically valid reorderings, insertions, or deletions of operators.
Because rewrite rules apply to combinations of operators, and because the different independently developed and maintained packages available for \stratosphere already contain more than 60 individual operators, it is practically impossible to define all rewrite rules across the different packages one-by-one. Instead, we define a concise set of rewrite templates using \ontology relationships and abstract operators as building blocks. Reasoning along relationships allows \optimizer to automatically instantiate the templates with concrete operators and thus enables us to derive individual rewrite options for concrete operator combinations on-the-fly. Currently, \optimizer requires only 10 rewrite templates, which are expanded to over 150 individual rewrite rules.

Figure~\ref{fig:rules} displays a subset of the available templates in Datalog notation; further rules cover different reorderings based on algebraic properties as well as insertion and removal of operators (not shown here for brevity).
The first three templates are static and can be evaluated at package loading time, whereas the last two templates are dynamic and are evaluable at query compile time only.
The first template covers commutative operators and expresses that two consecutive appearances of operators $X$ annotated as commutative in \ontology can be safely reordered.
Specifically, the goal \texttt{reorder(X,X)} evaluates to true if \ontology contains a \textit{hasProperty}-relationship of $X$ with the property \textit{commutative}.
Note that the commutativity does not necessarily need to be defined directly on $X$; the rule also applies if some ancestor of $X$ in \ontology is marked as commutative.
The second template (Line 4) enables reordering of operators based on the \textit{isA} relation and states that for any three operator instantiations $X,Y,Z$, the operators $X,Y$ are reorderable given that $X$ is not a prerequisite of $Y$, $X$ is a specialization of $Z$, and $Y,Z$ are reorderable.
We include the goal \texttt{not hasPrerequisite(Y,X)} in the templates to ensure that operator precedences are respected.
The third template (Lines 6--7) enables reorderings of consecutive $anntt$ operators $X,Y$, 
when $X$ is not a prerequisite of $Y$.

Dynamic rewrite templates are partly based on information not available before the query is compiled, for example, information on concrete attribute access by operators is only available after posing a query.
Template~4 (Lines 10--12) enables reordering of two single-input record-at-a-time operators if these operators have no read/write conflicts. This single rule essentially covers most optimization options achieved by~\cite{hueske_opening_2012}, which shows the power of our approach.

While most rules in \ontology are generic and apply to many operator combinations, other rules are more specific.
Suppose, we are given a dataflow that consists of an equi-join of two data sources $I_1,I_2$ followed by \textit{trnsf} that transforms only attributes of $I_1$, which are not part of the join condition.
This dataflow can be rewritten into an equivalent dataflow, which first applies \textit{trnsf} to $I_1$ and afterwards joins $I_1$ and $I_2$:

\noindent\begin{tikzpicture}[grow=right]
\tikzstyle{every node}=[font=\footnotesize]
\node[](O){$O$};
\node[rectangle,left=10pt of O](join){\textit{join}};
\node[left=10pt of join](trnsf){\textit{trnsf}};
\node[left=10pt of trnsf](A){$I_1$};
\node[below=0pt of trnsf](B){$I_2$};
\draw[-latex'](join)--(O);
\draw[-latex'](A)--(trnsf);
\draw[-latex'](B)--(join);
\draw[-latex'](trnsf)--(join);

\node[below left=0pt and 0pt of A](arrow){$\Leftrightarrow$};

\node[above left=1pt and 0pt of arrow](O1){$O$};
\node[left=10pt of O1](trnsf1){\textit{trnsf}};
\node[left=10pt of trnsf1](join1){\textit{join}};
\node[left=10pt of join1](A1){$I_1$};
\node[below=0pt of A1](B1){$I_2$};
\draw[-latex'](trnsf1)--(O1);
\draw[-latex'](join1)--(trnsf1);
\draw[-latex'](A1)--(join1);
\draw[-latex'](B1)--(join1);
\end{tikzpicture}

Similar to extending \ontology with new operators and properties, package developers can also extend the set of rewrite templates to enable dataflow optimization for their concrete application domain.
For example, the third template was added by the IE package developer, since it enables reordering \textit{anntt} instantiations, which is not supported by any other \ontology template.

\begin{figure}
\begin{Verbatim}[fontsize=\scriptsize,frame=single,numbers=left,numbersep=3pt,commandchars=\\\{\}]
\PY{c+c1}{\PYZpc{}\PYZpc{} static rules (package loading time) \PYZpc{}\PYZpc{}}
\PY{n+nf}{reorder}\PY{p}{(}\PY{n+nv}{X}\PY{p}{,}\PY{n+nv}{X}\PY{p}{)}\PY{p}{:-}\PY{n+nf}{hasProperty}\PY{p}{(}\PY{n+nv}{X}\PY{p}{,}\PY{l+s+sAtom}{'commutative'}\PY{p}{)}\PY{p}{.}\PY{c+c1}{\PYZpc{} 1}

\PY{n+nf}{reorder}\PY{p}{(}\PY{n+nv}{X}\PY{p}{,}\PY{n+nv}{Y}\PY{p}{)}\PY{p}{:-}\PY{o}{not} \PY{n+nf}{hasPrerequisite}\PY{p}{(}\PY{n+nv}{Y}\PY{p}{,}\PY{n+nv}{X}\PY{p}{)}\PY{p}{,}\PY{n+nf}{isA}\PY{p}{(}\PY{n+nv}{X}\PY{p}{,}\PY{n+nv}{Z}\PY{p}{)}\PY{p}{,}\PY{n+nf}{reorder}\PY{p}{(}\PY{n+nv}{Z}\PY{p}{,}\PY{n+nv}{Y}\PY{p}{)}\PY{p}{.}\PY{c+c1}{\PYZpc{}2}

\PY{n+nf}{reorder}\PY{p}{(}\PY{n+nv}{X}\PY{p}{,}\PY{n+nv}{Y}\PY{p}{)}\PY{p}{:-}\PY{n+nv}{isA}\PY{p}{(}\PY{n+nv}{X}\PY{p}{,}\PY{l+s+sAtom}{'anntt'}\PY{p}{)}\PY{p}{,}\PY{n+nv}{isA}\PY{p}{(}\PY{n+nv}{Y}\PY{p}{,}\PY{l+s+sAtom}{'anntt'}\PY{p}{)}\PY{p}{,}
              \PY{o}{not} \PY{n+nf}{hasPrerequisite}\PY{p}{(}\PY{n+nv}{Y}\PY{p}{,}\PY{n+nv}{X}\PY{p}{)}\PY{p}{.} \PY{c+c1}{\PYZpc{} 3}

\PY{c+c1}{\PYZpc{}\PYZpc{} dynamic rules (query compile time) \PYZpc{}\PYZpc{}}
\PY{n+nf}{reorder}\PY{p}{(}\PY{n+nv}{X}\PY{p}{,}\PY{n+nv}{Y}\PY{p}{)}\PY{p}{:-}\PY{n+nf}{hasProperty}\PY{p}{(}\PY{n+nv}{X}\PY{p}{,}\PY{l+s+sAtom}{'single-in'}\PY{p}{)}\PY{p}{,}\PY{n+nf}{hasProperty}\PY{p}{(}\PY{n+nv}{X}\PY{p}{,}\PY{l+s+sAtom}{'RAAT'}\PY{p}{)}\PY{p}{,}
              \PY{n+nf}{hasProperty}\PY{p}{(}\PY{n+nv}{Y}\PY{p}{,}\PY{l+s+sAtom}{'single-in'}\PY{p}{)}\PY{p}{,}\PY{n+nf}{hasProperty}\PY{p}{(}\PY{n+nv}{Y}\PY{p}{,}\PY{l+s+sAtom}{'RAAT'}\PY{p}{)}\PY{p}{,}
              \PY{o}{not} \PY{n+nf}{readWriteConflicts}\PY{p}{(}\PY{n+nv}{X}\PY{p}{,}\PY{n+nv}{Y}\PY{p}{)}\PY{p}{.} \PY{c+c1}{\PYZpc{} 4}

\PY{n+nf}{reorder}\PY{p}{(}\PY{n+nv}{X}\PY{p}{,}\PY{n+nv}{Y}\PY{p}{)}\PY{p}{:-}\PY{n+nf}{hasProperty}\PY{p}{(}\PY{n+nv}{X}\PY{p}{,}\PY{l+s+sAtom}{'single-in'}\PY{p}{)}\PY{p}{,}
              \PY{n+nf}{hasProperty}\PY{p}{(}\PY{n+nv}{X}\PY{p}{,}\PY{l+s+sAtom}{'|I|=|O|'}\PY{p}{)}\PY{p}{,}
              \PY{n+nf}{hasProperty}\PY{p}{(}\PY{n+nv}{X}\PY{p}{,} \PY{l+s+sAtom}{'S\PYZus{}in contains S\PYZus{}out'}\PY{p}{)}\PY{p}{,}
              \PY{n+nf}{hasProperty}\PY{p}{(}\PY{n+nv}{X}\PY{p}{,} \PY{l+s+sAtom}{'no field updates'}\PY{p}{)}\PY{p}{,}
              \PY{n+nf}{hasProperty}\PY{p}{(}\PY{n+nv}{Y}\PY{p}{,}\PY{l+s+sAtom}{'single-in'}\PY{p}{)}\PY{p}{,}
              \PY{n+nf}{hasProperty}\PY{p}{(}\PY{n+nv}{Y}\PY{p}{,}\PY{l+s+sAtom}{'|I|>=|O|}\PY{p}{)}\PY{p}{,}
              \PY{n+nf}{hasProperty}\PY{p}{(}\PY{n+nv}{Y}\PY{p}{,}\PY{l+s+sAtom}{'S\PYZus{}in = S\PYZus{}out'}\PY{p}{)}\PY{p}{,}
              \PY{o}{not} \PY{n+nv}{hasPrerequisite}\PY{p}{(}\PY{n+nv}{Y}\PY{p}{,}\PY{n+nv}{X}\PY{p}{)}\PY{p}{,}\PY{n+nf}{accessedFields}\PY{p}{(}\PY{n+nv}{Y}\PY{p}{,} \PY{n+nv}{ACCY}\PY{p}{)}\PY{p}{,}
              \PY{n+nv}{S\PYZus{}out}\PY{p}{(}\PY{n+nv}{X}\PY{p}{,}\PY{n+nv}{OUTX}\PY{p}{)}\PY{p}{,} \PY{n+nf}{contains} \PY{p}{(}\PY{n+nv}{OUTX}\PY{p}{,} \PY{n+nv}{ACCY}\PY{p}{)}\PY{p}{.} \PY{c+c1}{\PYZpc{} 5}
\end{Verbatim}
\vspace{-.5cm}
\caption{Rewrite template examples.}
\label{fig:rules}
\end{figure}

\subsection{Pay-as-you-go annotation of operators}
A key feature of \optimizer is its extensible design. Consider a new package for web analytics to be integrated into \stratosphere, containing an operator \textit{rmark} for detecting and removing HTML markup in web pages.
Initially, this operator would probably not be equipped with any \ontology annotations.
In this case, the \optimizer optimizer can infer only automatically detectable properties, i.e., reordering can only be performed on the basis of read/write-set analysis.
Later, the package developer invests some thought and annotates that \textit{rmark} outputs as many records as incoming ($|I|=|O|$). \optimizer infers from the set of automatically detectable properties, that \textit{rmark} is a single-input operator implemented with a map function and does not change the schema of the incoming records.
Taken these properties together, the last template of Figure~\ref{fig:rules} becomes applicable to \textit{rmark}.
A full specification of \textit{rmark} would include the definition of \textit{isA} relationships to other operators.
Actually, \textit{rmark} has the same semantics as the \textit{trnsf} operator from the Base package, as it essentially performs a transformation of the input texts.
Now all templates valid for \textit{trnsf} become applicable, such as the rule for reordering a \textit{join} and a \textit{trnsf} operator introduced in Section 4.2.
Given that \textit{rmark} accesses only attributes present in input $I_1$ that are not part of the join condition, \optimizer can then reorder a dataflow containing \textit{rmark} and \textit{join} as follows:

\noindent\begin{tikzpicture}[grow=right]
\tikzstyle{every node}=[font=\footnotesize]
\node[](O){$O$};
\node[rectangle,left=10pt of O](join){\textit{join}};
\node[left=10pt of join](rmark){\textit{rmark}};
\node[left=10pt of rmark](A){$I_1$};
\node[below=0pt of rmark](B){$I_2$};
\draw[-latex'](join)--(O);
\draw[-latex'](A)--(rmark);
\draw[-latex'](B)--(join);
\draw[-latex'](rmark)--(join);

\node[below left=0pt and 0pt of A](arrow){$\Leftrightarrow$};

\node[above left=1pt and 0pt of arrow](O1){$O$};
\node[left=10pt of O1](rmark1){\textit{rmark}};
\node[left=10pt of rmark1](join1){\textit{join}};
\node[left=10pt of join1](A1){$I_1$};
\node[below=0pt of A1](B1){$I_2$};
\draw[-latex'](rmark1)--(O1);
\draw[-latex'](join1)--(rmark1);
\draw[-latex'](A1)--(join1);
\draw[-latex'](B1)--(join1);

\end{tikzpicture}

\section{Algorithms}\label{sec:algorithms}
Given a dataflow $D$, the \optimizer optimizer performs two passes of the following three steps.
First, $D$ is analyzed for precedence relationships between operators based on rewrite templates and operator properties contained in \ontology.
This analysis yields a precedence graph, which is used in the plan enumeration phase, to secondly enumerate and thirdly rank valid plan alternatives based on a cost model.
Afterwards, the complex operators contained in $D$ are resolved into their elementary components and the three steps are repeated. Finally, the best plan is selected, translated, and physically optimized for parallel execution by the underlying execution engine (see~\cite{battre_nephele_2010} for details on this step). In the following, we discuss Phases 2--4 in more detail.

\subsection{Precedence analysis}\label{subsec:precedence}
Presto models dependencies between operators either explicitly on the basis of the \textit{hasPrecedence} relation or inferred, if the goal \texttt{reorder(X,Y)} fails for two operator instantiations $X,Y$.
The precedence graph construction starts by creating the directed transitive closure of the given dataflow, which explicitly models all pairwise operator execution orders in $D$.
The algorithm then inspects this graph to detect and remove edges that are not logically required.
It retains all edges incident to a data source or a sink to prevent reordering of sources and sinks.
The goal \texttt{reorder(X,Y)} is instantiated with start and end node of each edge $(u,v)$ and the inference mechanism tries to resolve the goal based on the operator properties and rewrite templates stored in \ontology. If  successful, both nodes are reorderable and the edge $(u,v)$ is removed from the precedence graph.
Precedence analysis is a polynomial time algorithm; its complexity is determined by computing the transitive closure in $O(|V|^3)$ using the Floyd-Warshall algorithm and the data complexity of stratified non-recursive Datalog, which we use for reasoning in \ontology~\cite{dantsin_complexity_2001}.

\begin{figure}
\centering
\includegraphics[width=\linewidth]{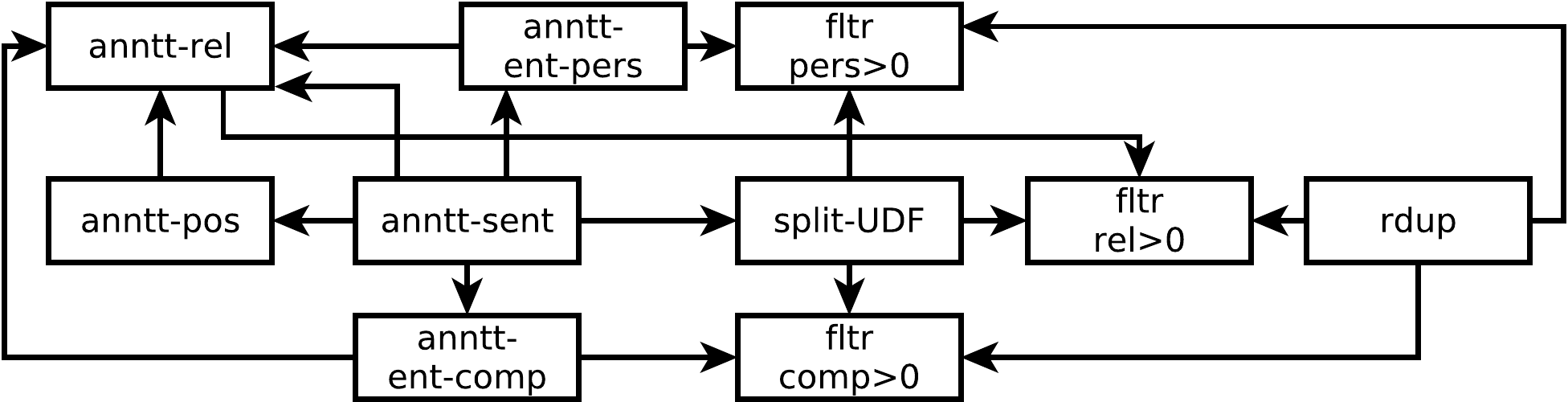}
\caption{Precedence graph for running example with complex operator resolution.}
\label{fig:precedence_example}
\end{figure}

Figure~\ref{fig:precedence_example} shows the final precedence graph for our running example (omitting data sources and sinks for readability). The displayed graph reflects precedences between DC, IE, and Base operators, e.g., \textit{rdup} and \textit{anntt-ent-person} are a prerequisite for the $\mathit{fltr_{person>0}}$ operator, and \textit{anntt-rel} is in a \textit{hasPrerequisite} relation with \textit{anntt-pos} (cf.\ Figure~\ref{fig:presto}(d)). The graph contains edges between \textit{anntt} and \textit{fltr} reflecting that the concrete instantiations of \textit{fltr} have read/write conflicts with their preceding \textit{anntt} operators.

\subsection{Plan enumeration}\label{subsec:enumeration}
Plan enumeration essentially generates different topological orders given by the precedence graph, while performing cost-based pruning. On contrast to topological sorting, the outcome are not full orders but DAG-shaped plans.
The main idea is to construct alternative plans for a given dataflow $D$ iteratively by analyzing the corresponding precedence graph for operators that have no outgoing edges.
Such operators are not required by any other operator and can therefore be added to the emerging physical dataflows.
If multiple operators have no outgoing edges, the algorithm creates a set of alternative partial plans.
The algorithm continues to pursue each alternative, removing the newly added operator from the precedence graph, estimating the costs of the partial plan (see Section~\ref{subsec:costs}), and pruning costly partial plan alternatives where possible.

\begin{figure}[t]
\centering
\includegraphics[width=\linewidth]{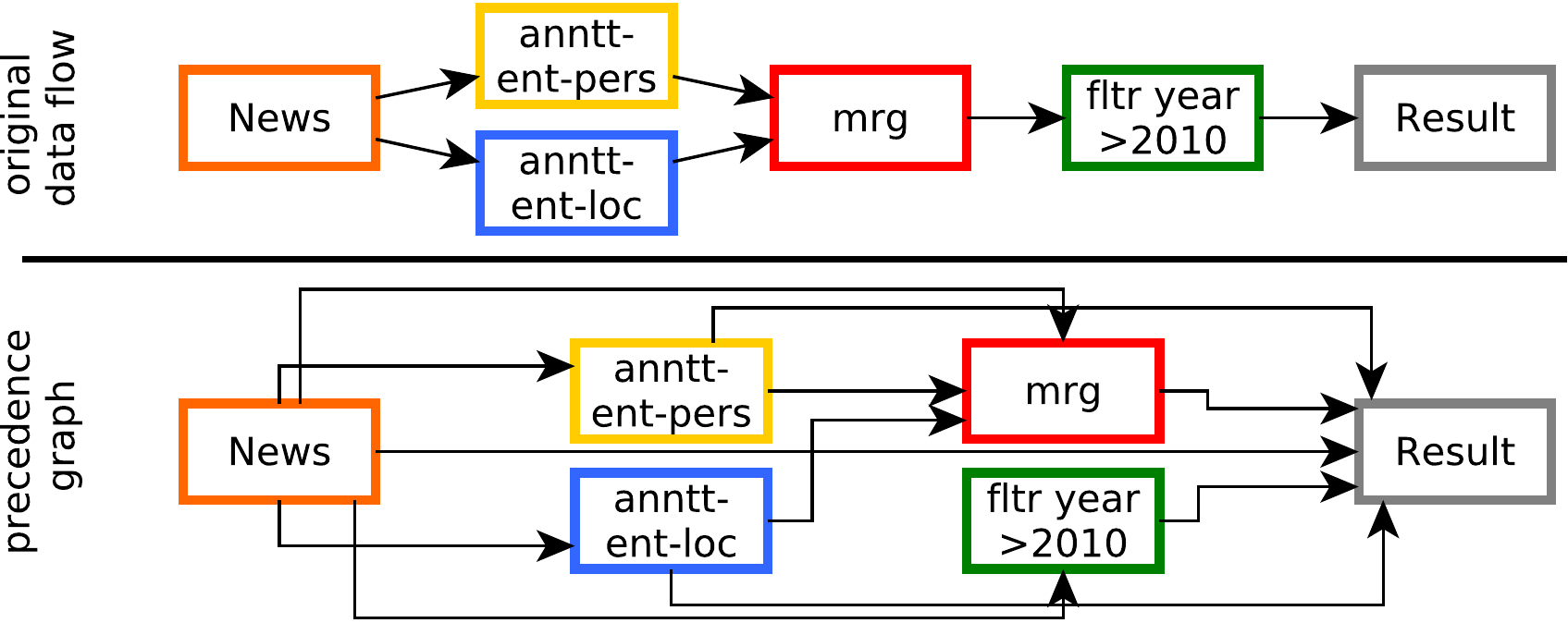}
\caption{DAG-shaped dataflow (top) and corresponding precedence graph (bottom) inspired by the running example.}
\label{fig:example_reordering}
\vspace{-1em}
\end{figure}

We explain its principles using the simplified dataflow shown in Figure~\ref{fig:example_reordering} (top).
Note that this dataflow is DAG-shaped, which poses no problem to \optimizer.
The dataflow performs task-parallel annotation of persons and companies.
Annotations are subsequently merged, and the result set is filtered for articles published after 2010.
The resulting precedence graph is displayed in Figure~\ref{fig:example_reordering} (bottom).

The recursive plan enumeration algorithm is displayed in Figure~\ref{fig:enumeration}.
It takes as input the original dataflow, the corresponding precedence graph, and a partial plan, which initially is  empty (Lines 1--2).
First, the algorithm selects the set of nodes from the precedence graph that have out-degree~0 (Line 8).
These operators are not a prerequisite of any remaining operator and can thus be added to the partial plan (Lines 11--13) without violating precedence constraints.
In our example, only the data sink can be selected.
Once added, the selected node is removed from the precedence graph (Line 14).
Since the partial plan was empty before adding the data sink, we cannot insert any edges in the partial plan and therefore, plan enumeration is recursively invoked again (Lines 16--17).
Now, \textit{mrg} and \textit{fltr} both have no outgoing edges anymore and are therefore added to the set of candidate nodes.
Each candidate node is processed individually, added to the partial plan and removed from the precedence graph. This yields in two alternative partial plans, which are both inspected further.

We exemplarily follow the plan with the \textit{mrg} operator.
The operator is added to the plan and subsequently, the set of $inputNodes$ is divided into required and optional nodes (Lines 19--24).
Required nodes are those nodes that have the currently added node as its direct predecessor in the original dataflow, optional successors are all other operators contained in $inputNodes$.
In our example, the set of required nodes is empty, and the set of optional nodes contains \textit{fltr}.
For each required node $m$, we create an edge $(n,m)$ for the newly added node $n$, add it to the edge set of our partial plan, estimate the costs of the partial plan, and
recursively call the plan enumeration algorithm (Lines 25--29).
Each optional node $l$ is processed individually.
We iteratively create edges $(n,l)$, estimate the costs of the new partial plan, and again recursively call the plan enumeration algorithm if necessary (Lines 30--34).
A recursive invocation of the plan enumeration algorithm terminates either if the precedence graph is empty and an alternative plan has been found (Lines 4--7), or if no alternative plans with smaller costs compared to the  initial plan were found (Line 33).

\begin{figure}[t]

\centering
\begin{Verbatim}[fontsize=\scriptsize,frame=single,numbers=left,numbersep=3pt,commandchars=\\\{\}]
\PY{n+nf}{enumAlternatives}\PY{o}{(}\PY{n}{Graph} \PY{n}{precedGraph}\PY{o}{,} \PY{n}{Graph} \PY{n}{plan}\PY{o}{,}
                      \PY{n}{Graph} \PY{n}{partialPlan}\PY{o}{)}\PY{o}{ \PYZob{}}

  \PY{k}{if} \PY{o}{(}\PY{n+na}{isEmpty}\PY{o}{(}\PY{n}{precedGraph}\PY{o}{)}{)} \PY{o}{\PYZob{}}
    \PY{n+na}{addPlanToResultSet}\PY{o}{(}\PY{n}{partialPlan}\PY{o}{)}\PY{o}{;}
    \PY{k}{return}{;}
  \PY{o}{\PYZcb{}}
  \PY{n}{candidateNodes} \PY{o}{=} \PY{n+na}{getNodesWithOutDegreeZero}\PY{o}{(}\PY{n}{precedGraph}\PY{o}{)}\PY{o}{;}

  \PY{k}{foreach}\PY{o}{(}\PY{n}{Node n} \PY{o}{in} \PY{n}{candidateNodes}\PY{o}{)} \PY{o}{\PYZob{}}
    \PY{n}{inputNodes} \PY{o}{=} \PY{n+na}{getNodesWithOpenInputs}\PY{o}{(}\PY{n}{partialPlan}\PY{o}{)}\PY{o}{;}

    \PY{n+na}{addNodeToPartialPlan}\PY{o}{(}\PY{n}{n}\PY{o}{)}\PY{o}{;}
    \PY{n+na}{removeNodeAndIncidentEdgesFromPrecedenceGraph}\PY{o}{(}\PY{n}{n}\PY{o}{)}\PY{o}{;}

    \PY{k}{if} \PY{o}{(}\PY{n+na}{isEmpty}\PY{o}{(}\PY{n}{inputNodes}\PY{o}{))}
      \PY{n+na}{enumAlternatives}\PY{o}{(}\PY{n}{precedGraph}\PY{o}{,} \PY{n}{plan}\PY{o}{,} \PY{n}{partialPlan}\PY{o}{)}\PY{o}{;}

    \PY{k}{foreach}\PY{o}{(}\PY{n}{Node m} \PY{o}{in} \PY{n}{inputNodes}\PY{o}{)}\PY{o}{\PYZob{}}
      \PY{c+cm}{/*split inputNodes into required and optional successors*/}
      \PY{k}{if} \PY{o}{(}\PY{n+na}{inputGraphcontainsEdge}\PY{o}{(}\PY{n}{n}\PY{o}{,}\PY{n}{m}\PY{o}{))}
        \PY{n+na}{addNodeToRequiredNodes}\PY{o}{(}\PY{n}{m}\PY{o}{)}\PY{o}{;}
      \PY{k}{else} \PY{n+na}{addNodeToOptionalNodes}\PY{o}{(}\PY{n}{m}\PY{o}{)}\PY{o}{;}
    \PY{o}{\PYZcb{}}
    \PY{k}{if}\PY{o}{(}\PY{o}{not} \PY{n+na}{isEmpty}\PY{o}{(}\PY{n}{requiredNodes}\PY{o}{)}\PY{o}{)} \PY{o}{\PYZob{}}
      \PY{n+na}{addEdgesToAllRequiredNodesInPartialPlan}\PY{o}{(}\PY{n}{m}\PY{o}{)}\PY{o}{;}
      \PY{k}{if} \PY{o}{(}\PY{n+na}{costs}\PY{o}{(}\PY{n}{partialPlan}\PY{o}{)} \PY{o}{<} \PY{n+na}{costs}\PY{o}{(}\PY{n}{originalPlan}\PY{o}{)}\PY{o}{)}
        \PY{n+na}{enumAlternatives}\PY{o}{(}\PY{n}{precedGraph}\PY{o}{,} \PY{n}{plan}\PY{o}{,} \PY{n}{partialPlan}\PY{o}{)}\PY{o}{;}
    \PY{o}{\PYZcb{}}
    \PY{k}{foreach}\PY{o}{(}\PY{n}{Node l} \PY{o}{in} \PY{n}{optionalNodes}\PY{o}{)} \PY{o}{\PYZob{}}
      \PY{n+na}{addEdgeToPartialPlan}\PY{o}{(}\PY{n}{n}\PY{o}{,}\PY{n}{l}\PY{o}{)}\PY{o}{;}
      \PY{k}{if} \PY{o}{(}\PY{n+na}{costs}\PY{o}{(}\PY{n}{partialPlan}\PY{o}{)} \PY{o}{<} \PY{n+na}{costs}\PY{o}{(}\PY{n}{originalPlan}\PY{o}{)}\PY{o}{)}
        \PY{n+na}{enumAlternatives}\PY{o}{(}\PY{n}{precedGraph}\PY{o}{,} \PY{n}{plan}\PY{o}{,} \PY{n}{partialPlan}\PY{o}{)}\PY{o}{;}
    \PY{o}{\PYZcb{}}
  \PY{o}{\PYZcb{}}
  \PY{n+na}{addPlanToResultSet}\PY{o}{(}\PY{n}{plan}\PY{o}{)}\PY{o}{;}
  \PY{k}{return}\PY{o}{;}
\PY{o}{\PYZcb{}}
\end{Verbatim}
\vspace{-.5cm}
\caption{Plan enumeration with SOFA.}
\label{fig:enumeration}
\end{figure}

\begin{figure*}[t]
\centering
\includegraphics[width=\linewidth]{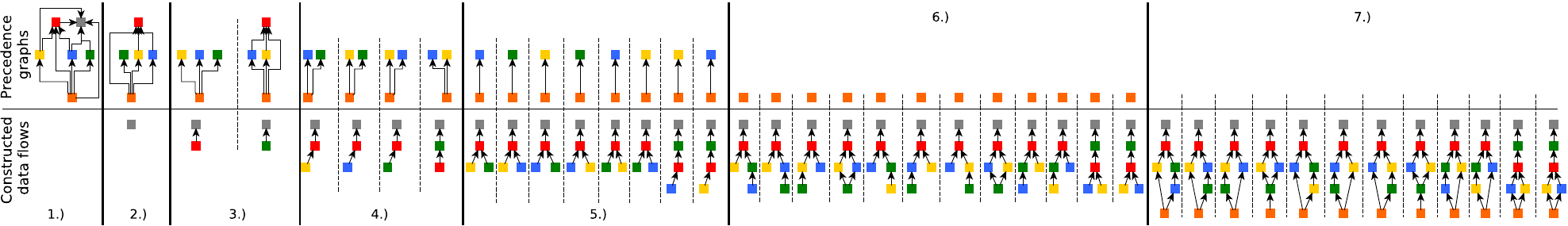}
\caption{Plan enumeration for the DAG-shaped dataflow from Figure~\ref{fig:example_reordering}. Columns are alternative partial plans grouped into stages of the algorithm. Boxes correspond to operators with isochromatic frames as defined in Figure~\ref{fig:example_reordering}. Stage~7 contains all valid plans for this dataflow.}
\label{fig:plan_enumeration_example}
\end{figure*}

Figure~\ref{fig:plan_enumeration_example} shows all stages of enumerating the plan space for the dataflow from Figure~\ref{fig:example_reordering}.
In Stage~1, only the data sink can be selected (grey box) and is thus removed in Stage~2.
The algorithm can now either add \textit{mrg} or \textit{fltr} (red and green boxes) as both have no outgoing edges in Stage~3.
In Stage~4, the \textit{mrg} plan results in three further alternatives, namely \textit{fltr}, \textit{anntt-ent-comp}, and \textit{anntt-ent-pers}.
At that point, only the source cannot be added to the plan yet and consequently the remaining two operators can be added in arbitrary order.
Thus, the four alternatives are expanded to eight plans in Stage~5.
Finally, the last operator and source is added in Stage~6 and 7 resulting in 12 different alternatives.

\noindent\textbf{Pruning.} The plan enumeration algorithm has exponential worst-case complexity (consider for instance a precedence graph without any edges). We included a simple technique for search space pruning in our algorithm preventing completion of partial plans whose estimated costs are higher than the estimated costs for the current best dataflow. Once a cheaper plan was found, we update the costs of the best plan, in a manner similar to accumulated cost pruning in top-down query optimization~\cite{graefe_volcano_1994,graefe_cascades_1995}. If no alternative plan with lower estimated costs compared to the best plan could be constructed, we terminate (cf.\ Figure~\ref{fig:enumeration}, Line 33).

\subsection{Cost estimation}
\label{subsec:costs}
To estimate costs and result sizes of a dataflow, \optimizer depends on estimates for key figures of operators, which can either be provided by the developer by adding appropriate annotations to \ontology, by sampling from the input data, or by runtime monitoring of previously executed dataflows.
We estimate the costs of a plan by computing the weighted sum of estimated ship data volume, I/O volume, and CPU usage of the \udfs per stub call.

Specifically, the costs of an operator $o_i$ are estimated as follows: let $c_i$ be the average CPU usage of $o_i$ per invocation, $s_i$ the estimated startup costs of $o_i$, and $r_i$ the estimated number of processed input items of $o_i$.
Including startup times of operators is particularly important for complex non-relational \udfs, as many IE and DC operators need a long startup time for instance to load large dictionaries, or to assemble trained models.
Furthermore, let $d_i$ denote the estimated I/O costs of an input item processed by $o_i$, $n_i$ the estimated shipping costs of an output item produced by $o_i$, and $sel_i$ the selectivity of $o_i$.
The estimated number of items $r_i$ processed by an operator $o_i$ is calculated as $r_i = \sum_{(h,i)\in E(D)}r_h*sel_h$.
The costs of an operator $o_i$ are estimated as the weighted sum of the estimated ship data volume, I/O volume, and CPU usage of $o_i$ using the formula $costs(o_i)=w*(c_i*r_i+s_i)+u*(d_i*r_i)+v*(n_i*r_i*sel_i)$, where $u,v,w$ denote weight constants for each cost component.
Note that the formula for operator costs can be replaced with custom cost functions, which are added to \ontology by the package developers.
For example, to accurately estimate the costs for dataflows containing IE operators, we also capture the \textit{projectivity} of \textit{anntt} operators, i.e., the average number of annotations produced by an \textit{anntt} instantiation.
Consequently, the selectivity of $\mathit{fltr_{anntt}}$ is denoted as $sel(fltr_{anntt}) = r_{i-1}*proj(anntt)$.

Finally, the total costs of a dataflow $D$ are estimated as $costs(D) = \sum_{i=1}^n costs(o_i)$.
Note that our cost model optimizes for total computation time, disregarding parallelization in the underlying execution engine. However, we see in Section~\ref{sec:evaluation} that this approach already allows us to correctly rank enumerated plan alternatives in many cases.

\section{Related Work}
\label{sec:relatedwork}
Query optimization is a prominent topic in database research and many facets of our problem setting have been addressed before.
The optimizer of the Starburst project leverages an extensible set of rules to rewrite query expressions~\cite{haas_extensible_1989}.
Rules are specified as pairs of condition and action functions implemented in a procedural language~\cite{pirahesh1992extensible}.
Volcano~\cite{graefe_volcano_1994} and Cascades~\cite{graefe_cascades_1995} have extensible sets of transformation and implementation rules, which rewrite query expressions and compile them to physical execution plans.
Chaudhuri \etal proposed declarative rewrite rules to improve the optimization of queries with external functions~\cite{chaudhuri1993query}.
All aforementioned approaches deal with the optimization of relational queries and assume that rewrite rules are manually added to the optimizer framework.
Our work differs as it focuses on \udf-heavy query, also deals with DAG-shaped plans, and does not require the explicit definition of condition--action rules for new operators but automatically infers them from a concise set of operator properties.
This considerably increases extensibility. 

The optimization of relational queries with user-defined predicates has been another focus of research~\cite{hellerstein1993predmig, chaudhuri_optimization_1999, srivastava_query_2006}.
In these works, the semantics of the \udfs to be reordered is assumed to be uniform, \ie, filtering tuples.
Consequently, these approaches focus on the problem of identifying the optimal order of filters and do not address the question whether general \udfs can be reordered or not; besides, they disregard the effects of parallelization functions on \udfs.
In contrast, our work addresses the optimization of Map/Reduce-style dataflows with user-defined operators for which additional information is required to answer this question.
In our approach, this information is provided by the developer or inferred from \ontology.

While common query optimizers use tree-shaped query execution plans, our work focuses on DAG-shaped dataflows in the spirit of~\cite{neumann2005dagplans}.
There, Neumann investigated the problem of generating and executing DAG-shaped query evaluation plans for relational queries in order to enable sharing of intermediate results.
This is different from our setting as the input of our optimizer are already DAG-shaped dataflows in contrast to declarative relational queries.

Several methods to optimize more general dataflows have been proposed.
Ogasawara \etal propose an algebraic approach to define scientific workflows~\cite{ogasawara2011algsciflow} and optimize their execution.
Operators are classified as \udfs with strict templates or as relational expressions.
While their classification of \udfs is somewhat similar to particular combinations of our operator properties, we regard our approach as more general as it
features a richer set of properties that can be freely composed to precisely reflect the characteristics of an operator.
Furthermore, their work
does not contain a transformation-based optimizer.
Simitsis \etal present an approach for optimizing ETL processes~\cite{simitsis_optimizing_2005}. They introduce three
different optimization techniques, namely reorderings of adjacent single-input/single-output operators that have no
read/write-conflicts, merging and splitting of operators, and factorization of operators. Our approach is more general
as \optimizer is able to optimize arbitrary dataflows and it is able to reorder any operator combinations given that
precedence constraints are respected.
Stubby is an optimizer for workflows constructed of multiple Map/Reduce jobs~\cite{lim2012stubby}.
This approach is orthogonal to \optimizer, as it leverages manual annotations of Map and Reduce functions to merge Map/Reduce jobs while preserving the logical order of operations.
The MRQL framework performs physical algebraic optimizations of Map/Reduce dataflows consisting only of
relational operators~\cite{fegaras_mrql_2012}, but disregard general \udfs. Both approaches identify valid transformations from static rules and
operator properties, such as in- and output schema. Our approach differs in using an extensible taxonomy of a
richer set of operator properties and rewrite rules to deduce precedence constraints.
Work presented by Hueske \etal~\cite{hueske_opening_2012} on reordering operators in Map/Re\-duce-style dataflows is based on specific information about the operator's behavior in terms of accessed data attributes obtained by static code
analysis.
We included this idea in \optimizer, but also show how semantic annotations that
describe the behavior of \udfs more precisely can help to further increase the space of possible rewritings; besides, our
approach allows to rewrite DAG-shaped dataflows, which is not possible with the algorithm presented
in~\cite{hueske_opening_2012}. The SUDO optimizer~\cite{zhang_optimizing_2012} combines manual annotation and code analysis to analyze \udf properties with respect to data partitioning to avoid unnecessary data shufflings.
This problem is orthogonal to \optimizer, where we analyze semantic operator properties to reduce execution times by
reordering operators. Pig Latin~\cite{olston2008pig} is a procedural higher-level language for Hadoop and features a ``safe'' optimizer. It applies a limited set of heuristic transformation rules, such as filter push down, that most likely is beneficial and relies otherwise on the decisions of the programmer.

In summary, we believe that \optimizer is the first extensible, fully functional optimizer for arbitrary DAG-shaped dataflows for Map/Reduce-style systems.

\begin{table*}[!ht]
\centering
\begin{tabular}{|c|c|c|c|c|c|c|c|}
\hline
&Q1&Q2&Q3&Q4&Q5&Q6&Q7\\
\hline
\optimizer&\textbf{4545 (783)}&\textbf{5 (5)}&\textbf{7624 (844)}&\textbf{12 (10)}&\textbf{6 (4)}&\textbf{4
(4)}&\textbf{4 (2)}\\
Hueske \etal~\cite{hueske_opening_2012}&512 (214)&1 (1)&\textbf{7624 (844)}&1 (1)&1 (1)&\textbf{4 (4)}&1 (1)\\
Olston \etal~\cite{olston2008pig}&1 (1)&1 (1)&240 (192)&\textbf{6 (6)}&1 (1)&2 (2)&1 (1)\\
Simitis \etal~\cite{simitsis_optimizing_2005}&38 (38)&1 (1)&240 (192)&4 (4)&2 (2)&2 (2)&1 (1)\\
\hline
\end{tabular}
 \caption{Number of plan alternatives per query.
Counts in braces denote the number of plans considered with pruning enabled.
Bold numbers indicate the plan space containing the fastest measured plan (see Figure~\ref{fig:competitors}).}
 \label{tab:planAlternatives}
\end{table*}

\section{Evaluation}
\label{sec:evaluation}
We evaluated \optimizer on a 28-node cluster, each equipped with a 6-core Intel Xeon E5
processor, 24~GB RAM, and 1TB HDD using Stratosphere 0.2.1.

\noindent\textbf{Queries.}
We implemented, optimized, and executed seven \udf-heavy queries originating from different application
domains\footnote{Queries may be found at http://bit.ly/1dP9Nm2, datasets are available on request.}. Q1, Q2, and Q7 are
pipeline-shaped, Q3 and Q6 are tree-shaped, and Q4 and Q5 are DAG-shaped.

\textbf{Q1} adopts the dataflow described in our running example for relationship extraction from biomedical
literature using IE and DC \udfs.
\textbf{Q2} performs an advanced word count by computing term frequencies in a corpus grouped by year.
The query first splits the input data into sentences, reduces terms to their stem, removes stopwords, splits the text into tokens, and aggregates the token counts by year.
\textbf{Q3} extracts NASDAQ-listed companies that went bankrupt between 2010 and 2012 from a subset of Wikipedia.
This query takes article versions from two different points in time, annotates company names in both sets and applies
different \textit{fltr} operators and a \textit{join} to accomplish the task.
\textbf{Q4} corresponds to the dataflow shown in Figure~\ref{fig:example_reordering} and performs task-parallel
annotation of person and location names.
\textbf{Q5} analyzes DBpedia to retrieve politicians named 'Bush' and their corresponding parties using a mixture of DC
and base operators.
\textbf{Q6} is a relational query inspired by the TPC-H query 15.
It filters the lineitem table for a time range, joins it with the supplier table, groups the result by join key, and aggregates the total revenue to compute the final result.
\textbf{Q7} uses two complex IE operators to split incoming texts into sentences and to extract person names.

\noindent\textbf{Datasets.}
We evaluated Q1 on a set of 10 million randomly selected citations from Medline,
Q2 was evaluated on a set of 100,000 full-text articles from the English Wikipedia initially published between 2008 and 2012,
Q3 was evaluated on two sets of English Wikipedia articles of 50,000 articles each, one set from 2010 and one set from 2012,
Q4 and Q7 on a set of 100,000 full-text articles from the English Wikipedia downloaded in 2012,
Q5 on the full DBpedia dataset v.~3.8,
and Q6 was evaluated on a 100GB relational dataset generated using the TPC-H data generator.
For each experiment, we report the average of three runs.
Estimates on operator selectivities, projectivities, startup costs, and average execution times per input item were derived from 5\% random samples of each dataset.

\noindent\textbf{Competitors.}
Although dataflow languages for Big Data are a hot topic in current research, surprisingly few systems actually optimize the dataflow at the logical level as we do.
Thus, detecting appropriate competitors is difficult, because optimizers are commonly deeply coupled to a particular system.
We reimplemented the ideas of three current dataflow optimizers, namely techniques presented by Hueske
\etal~\cite{hueske_opening_2012}, Olston \etal~\cite{olston2008pig}, and Simitsis
\etal~\cite{simitsis_optimizing_2005}.
We compare the number of plan alternatives found and the achieved runtime improvements.
For each method, we disabled rules and information on operator properties stored in \ontology and replaced them with the
appropriate rewrite rules described in~\cite{hueske_opening_2012, olston2008pig, simitsis_optimizing_2005}.
For the method of Olston et al., we referred to the online documentation of rewrite rules for Apache Pig, version 0.11.1.
For Hueske et al., we enabled annotation of read- and write-sets, but disabled reordering of DAG-shaped plans.

\begin{figure*}[t]
  \centering
\includegraphics[width=1\linewidth]{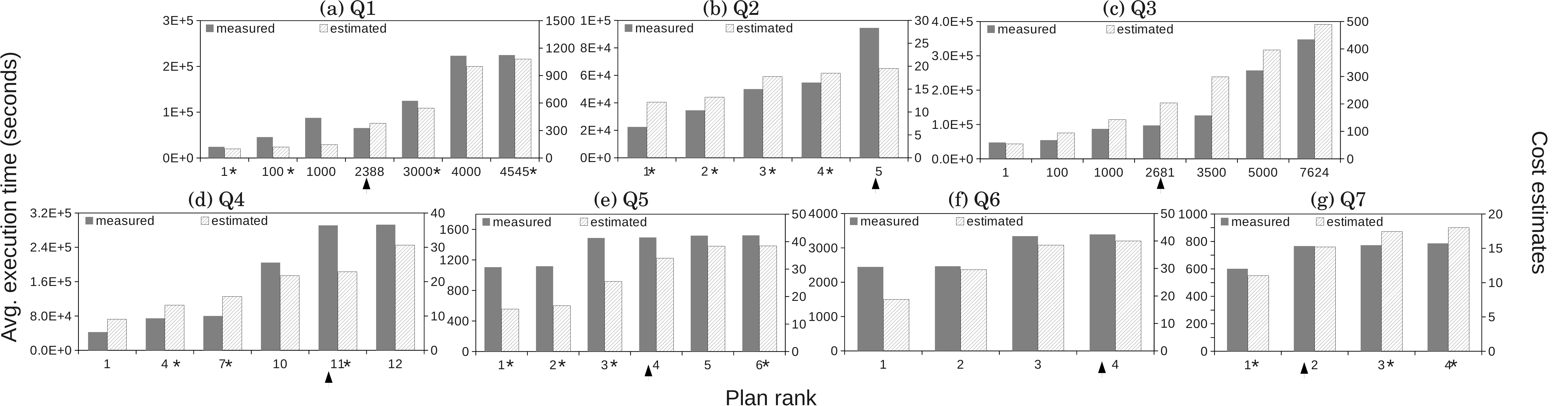}
  \caption{Cost estimates and execution time for queries. Ranks marked with a '*' denote
  plans found only with \optimizer. Arrows below each figure point to the time required by executing the query without any optimization.}
  \label{fig:runtimes}
\end{figure*}
\subsection{Finding optimal plans}
A large number of semantically equivalent plans for a concrete dataflow has the potential to contain the most effective variant.
Therefore, we first evaluate \optimizer to all three competitors with respect to the number of alternative plans found with each method.
We turned search space pruning off and enumerated the complete space of alternative dataflows for all queries.
In Section~\ref{sec:approach}, we explained how complex operators can be resolved into a series of interconnected elementary operators.
Q1, Q2, and Q7 contain complex operators, thus, we enumerated the plan space for these queries both using only elementary operators and using combinations of elementary and complex operators.
For the methods presented in~\cite{hueske_opening_2012,simitsis_optimizing_2005,olston2008pig}, we used complex operators only, as this methods do not provide mechanisms for operator expansion.

As displayed in Table~\ref{tab:planAlternatives}, \optimizer enumerates the largest plan space in all cases.
The method presented by Hueske \etal is unable to rewrite Q2, Q4, Q5, and Q7, because it is neither capable of rewriting DAG-shaped dataflows (Q4, Q5) nor of expanding complex operators (Q2, Q7).
The approach of Olston~\etal can rewrite only Q3, Q4, and Q6, because these are the only methods that involve filter push-ups.
Simitis~\etal find no alternative plans for Q2 and Q7, as in these cases, no adjacent single-input/single-output operators were reorderable.
For Q3 and Q6, \optimizer and~\cite{hueske_opening_2012} both enumerate the largest plan space, as for both queries the predominant rewrite options concerned \textit{fltr} operators.

To evaluate the correctness of plan ranking performed by \optimizer, we sorted the complete plan space for each query plan ascending by estimated costs, selected different plans from a rank interval, and report estimated costs and observed runtimes for these plans.
As shown in Figure~\ref{fig:runtimes}, \optimizer ranks the different algebraic execution plans correctly, and for Q1, Q2, Q5, and Q7, the best ranked plans were retrieved only with \optimizer.

We also observed a large optimization potential for most tasks.
For example, the best ranked plans for Q1--Q4 outperform the worst ranked plans with factors in the range of 4.2~(Q2) to 9.1~(Q1).
For the remaining queries Q5--Q7 we observed differences in execution times of 23 to 28 \% between the best and worst plan.
Note that these three queries were the shortest running in our experiments with total runtimes between 10 to 30 minutes, and a significant portion of these runtimes can be attributed to system initialization and communication.
Thus, we expect that these queries benefit much more from optimization on larger datasets.

\subsection{Pruning}
Table~\ref{tab:planAlternatives} displays the plan space with search space pruning enabled in brackets.
For queries spanning the largest plan space (Q1 and Q3), pruning helps to significantly reduce the enumerated plan space.
For the methods presented in~\cite{olston2008pig,simitsis_optimizing_2005}, which both enumerate significantly smaller plan spaces than \optimizer, pruning as performed by our enumeration algorithm does not reduce the plan space in most cases. 
For each tested query, the optimization time with pruning enabled takes not longer than 2.5~seconds with \optimizer. Enumerating the complete plan space for each query takes at most 10~seconds, which is negligible compared to the execution times of our long-running evaluation queries. Note that the largest part of these optimization times can be attributed to reasoning along \ontology relationships, which could be improved in many known ways~\cite{sagiv_optimizing_1987}.

\subsection{Optimization benefits}
\begin{figure*}[t]
  \centering
\includegraphics[width=1\linewidth]{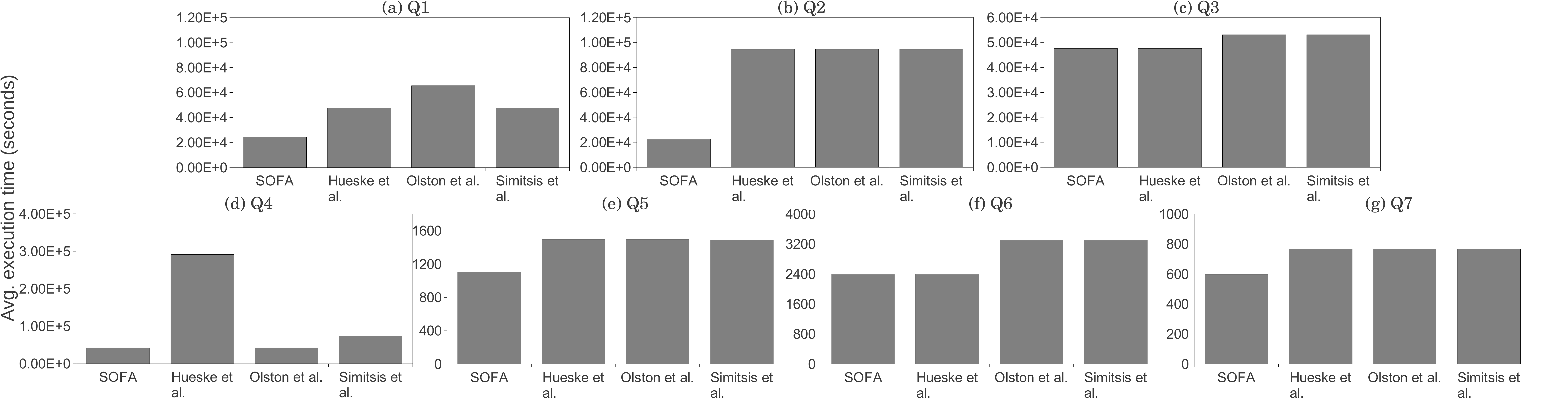}
  \caption{Execution times of best plans found with SOFA and best plans found by three competitors.}
  \label{fig:competitors}
\end{figure*}

In our third experiment, we evaluated to which extent dataflow optimization benefits from information on operator semantics. Figure~\ref{fig:competitors} displays the execution times of the best ranked plan found with \optimizer as well as the methods described in \cite{hueske_opening_2012,olston2008pig,simitsis_optimizing_2005}.
For each tested query, \optimizer finds the fastest plan, and for Q1, Q2, Q5, and Q7, \optimizer finds significantly faster plans than competitors: the best plan found with \optimizer outperforms the best plans found by~\cite{hueske_opening_2012} with factors of up to 6.8 (Q4), by~\cite{olston2008pig} and~\cite{simitsis_optimizing_2005} with factors up to 4.2 (Q2).
The method of Hueske et al. performs as well as \optimizer for Q3 and Q6, because both methods enumerate the same plan space for these two queries.
The rewrite rules of Olston et al. and Simitsis et al. find the same best plan as \optimizer for Q4.
In these cases, plan optimization involves only reordering filter operators, which is addressed equally well in these methods as in \optimizer.
Note that the method of Hueske et al.\ cannot rewrite Q4, as this query is DAG-shaped.
All other queries involve rewriting general \udfs and expansion of complex operators, and thus, optimization benefits notably from semantic information that is available in \optimizer.

\subsection{Extensibility}\label{subsec:evalextensibility}
Finally, we concretize the example from Section 4.3 to quantify the effect of pay-as-you-go annotation of operators in \optimizer. Recall the novel \textit{rmark} operator, which replaces HTML tags in web pages by a series of `\%' of the same length as the removed tags to retain text length and markup position.
Imagine a query Q8 that first replaces HTML markup in websites, computes term frequencies from the websites content, and finally filters terms starting with a series of `\%'.
The high-level dataflow looks as follows:

\noindent\begin{tikzpicture}[grow=right]
\tikzstyle{every node}=[font=\scriptsize]
\node[](O){$O$};
\node[rectangle,left=3pt of O](fltr){\textit{fltr}};
\node[left=3pt of fltr](grp){\textit{grp}};
\node[left=3pt of grp](splttok){\textit{splt-tok}};
\node[left=3pt of splttok](stop){\textit{rm-stop}};
\node[left=3pt of stop](stem){\textit{stem}};
\node[left=3pt of stem](spltsent){\textit{splt-sent}};
\node[left=3pt of spltsent](rmark){\textit{rmark}};
\node[left=3pt of rmark](in){$I$};

\draw[-latex'](fltr)--(O);
\draw[-latex'](grp)--(fltr);
\draw[-latex'](splttok)--(grp);
\draw[-latex'](stop)--(splttok);
\draw[-latex'](stem)--(stop);
\draw[-latex'](spltsent)--(stem);
\draw[-latex'](rmark)--(spltsent);
\draw[-latex'](in)--(rmark);
\end{tikzpicture}

Initially, $\textit{rmark}$ is annotated only with an \textit{isA}-relationship to the abstract \ontology concept \textit{operator}.
In this case, \optimizer can analyze only read and write access on attributes similar to the method presented in~\cite{hueske_opening_2012}, which yields in 10 semantically equivalent plans for Q8.
After adding the information that \textit{rmark} is a schema preserving record-at-a-time operator implemented with a Map function, \ontology already finds 18 equivalent algebraic plans.
Finally, when $\textit{rmark}$ is fully specified, including an \textit{isA}  relationship to the Base operator \textit{trnsf}, \optimizer would find 75 alternative plans.

\section{Conclusions}
\label{sec:conclusions}

We addressed the problem of logical optimization for \udf-heavy dataflows and present \optimizer, a novel, extensible, and comprehensive optimizer. \optimizer builds on a concise set of properties describing the semantics of Map/Reduce-style \udfs  and a small set of rewrite templates to derive equivalent plans.
A unique characteristic of our approach is extensibility: we arrange operators and their properties into a taxonomy, which considerably eases integration and optimization of new operators. We implemented our solution in \stratosphere, a fully-functional system for large-scale data analytics. Our experiments reveal that \optimizer is able to reorder acyclic dataflows of arbitrary shape (pipeline, tree, DAG) from different application domains, leading to considerable runtime improvements. We also show that \optimizer finds plans that clearly outperform those from other techniques.

\balance

\section{Acknowledgments}
This research was funded by the German Research Foundation under grant ``FOR 1036: Stratosphere-Information Management
on the Cloud.'' We thank Martin Beckmann and Anja Kunkel for help with implementing the Meteor queries we used for 
evaluation, and we thank Volker Markl and Stephan Ewen for valuable discussion and feedback.

\bibliographystyle{abbrv}
\small
\bibliography{bibliography}
\end{document}